\documentclass[10pt,journal,compsoc]{IEEEtran}
\IEEEoverridecommandlockouts
\usepackage{float}
\usepackage{cite}
\usepackage{amsmath,amssymb,amsfonts}
\usepackage{algorithmic}
\usepackage{graphicx}
\usepackage{textcomp}
\usepackage{xcolor}
\usepackage{listings}
\usepackage{subcaption}
\usepackage[ruled,vlined]{algorithm2e}
\usepackage{url}

\def\BibTeX{{\rm B\kern-.05em{\sc i\kern-.025em b}\kern-.08em
    T\kern-.1667em\lower.7ex\hbox{E}\kern-.125emX}}
    
\begin{document}
\title{GPU Tensor Cores for fast Arithmetic Reductions 
\thanks{}
}

\author{Crist\'obal A. Navarro, Roberto Carrasco, Ricardo J. Barrientos, Javier A. Riquelme and Raimundo Vega
\IEEEcompsocitemizethanks{
\IEEEcompsocthanksitem C. A. Navarro, R. Carrasco and R. Vega are with the Institute of Informatics of Universidad Austral de Chile.
\IEEEcompsocthanksitem R. Barrientos and J. Riquelme are from the 
Laboratory of Technological Research in Pattern Recognition (LITRP), Dept. DCI, Faculty of Engineering Science, Universidad Católica del Maule, Chile.
\protect\\
e-mail: cnavarro@inf.uach.cl}}

\onecolumn

\IEEEtitleabstractindextext{%
\begin{abstract}
This work proposes a GPU tensor core approach that encodes the arithmetic
reduction of $n$ numbers as a set of chained $m \times m$ matrix multiply accumulate
(MMA) operations executed in parallel by GPU tensor cores.
The asymptotic running time of the proposed chained tensor core approach is $T(n)=5 log_{m^2}{n}$ and its speedup is $S=\dfrac{4}{5} log_{2}{m^2}$ over the classic $O(n \log n)$ parallel reduction algorithm. Experimental performance results show that the proposed reduction method is $\sim 3.2 \times$ faster than a conventional GPU reduction implementation, and preserves the numerical precision because the sub-results of each chain of $R$ MMAs is kept as a 32-bit floating point value, before being all reduced into as a final 32-bit result. 
The chained MMA design allows a flexible configuration of thread-blocks; small thread-blocks of 32 or 128 threads can still achieve maximum performance using a chain of $R=4,5$ MMAs per block, while large thread-blocks work best with $R=1$. The results obtained in this work show that tensor cores can indeed provide a significant performance improvement to non-Machine Learning applications such as the arithmetic reduction, which is an integration tool for studying many scientific phenomena. 
\end{abstract}

}

\maketitle

\section{Introduction}
The recent rise of Deep Learning applications \cite{lecun2015deep,SCHMIDHUBER201585}  has made an impact not only in the software industry, but in the hardware 
industry as well, influencing chip manufacturers to include application specific integrated circuits (ASICs) that execute dedicated matrix operations to speedup training and inference in deep neural networks (DNN) \cite{jouppi2018motivation, tpu_google_2017}. 
Such is the case of Nvidia, who started including Tensor Cores (TC) inside GPU chips to further accelerate Deep Learning applications \cite{tensor_cores_2018, martineau2018benchmarking}. As of January 2020, GPUs contain up to 640 tensor cores that can work in parallel. Each tensor core is a hardware-implemented function that performs a matrix multiply accumulate (MMA) operation of $4 \times 4$ matrices in one GPU clock cycle, \textit{i.e.}, $D_{4\times 4} = A_{4\times 4} \times B_{4 \times 4} + C_{4 \times 4}$. Tensor cores play an important role
at leveraging the performance of real-time Deep Learning applications, such as autonomous driving, video-camera processing, real-time anomaly detection, as well as traditional linear algebra operations such as matrix multiplication of large matrices. 
Actual GPU tensor core performance can be up to an order of magnitude faster than regular FP32/INT32 GPU cores. However, adapting any arbitrary algorithm to a tensor core scheme is not a trivial task, as tensor cores are different from regular GPU cores. While GPU cores are capable of executing a whole instruction set (\textit{i.e.}, the instructions used in a regular CUDA/OpenCL program), tensor cores are capable of executing one operation but significantly faster; a matrix multiply accumulate (MMA) over $4\times 4$ matrices, in one GPU clock cycle. This fast but restrictive operation offered by tensor cores presents an interesting research opportunity as well as challenge to the HPC community when trying to adapt non-Machine Learning computational patterns to GPU tensor cores and exploit the potential performance available. One important non-Machine Learning computational pattern is the arithmetic reduction \cite{10.1007/BFb0016236}, which is one of the most used patterns in science and technology, \textit{i.e.}, it is the discrete integration tool for modelling many scientific phenomena, from n-body/Monte Carlo simulations \cite{bedorf2012sparse, NAVARRO201648}, cellular automata \cite{wolfram1983statistical} to map-reduce workloads \cite{stuart2011multi} and ray tracing \cite{gunther2007realtime}, among many others. 

This work improves the preliminary results of an earlier conference work \cite{8705253} and proposes an efficient way to exploit the tensor core performance of GPUs to accelerate the computation of the arithmetic reduction of $n$ numbers. The reduction algorithm is re-designed as a set of simultaneous chains of $m \times m$ matrix multiply accumulate (MMA) operations followed by a global reduction of all the partial MMA results, to form one final reduction value. The cost of reducing $n$ numbers, using a PRAM-like GPU computing model, becomes $T(n)=5 \log_{m^2}{n}$, giving a speedup of $S=(4/5) \log_{2}{m^2}$ over the traditional PRAM-CREW parallel reduction algorithm of running time $O(n \log n)$. The experimental results support the theoretical ones, showing up to $Q\times$ of speedup over a traditional parallel GPU reduction in Nvidia's CUB library. The results obtained in this work show that there is significant potential performance that can be exploited from tensor cores, and the reduction algorithm proposed in this work serves as a positive case where tensor cores can accelerate some non-Machine Learning computations.

The rest of the manuscript is organized as follows; an overview of GPU Tensor Core programming is presented in Section \ref{sec_tensor_cores} and related works are considered in Section \ref{sec_related_work}. The formulation of the new tensor core based reduction algorithm and its analysis are presented in Section \ref{sec_formulation_analysis}, and different implementation variations of the algorithm are described in Section \ref{sec_variants}, where the best one is chosen. An experimental performance comparison against other GPU reductions is presented 
in Section \ref{sec-comparison-cub}. Finally, a discussion of the results as well as conclusions are given in Section \ref{sec_discussion}.

\section{GPU Tensor Core Programming}
\label{sec_tensor_cores}
\textit{Note: the reader can skip this section if he/she is already familiar with GPU Tensor Core programming.}

GPU Computing \cite{navarro_hitschfeld-kahler_mateu_2014} has become a useful tool for reducing the computation time on large scale problems as well as for bringing real-time performance on many applications. Currently, GPU Computing has been used successfully in physics \cite{CARTER2018148, NAVARRO201648}, all-pairs problems \cite{8392762, Chan:2007:MAA:1250790.1250877}, medicine \cite{0031-9155-56-22-002}, image processing \cite{CERDA20188}, deep learning \cite{SCHMIDHUBER201585} and computer graphics \cite{Chaitanya2017}, among many other fields \cite{Haidar:2017}. One of the key aspects of GPU computing is its programming model which offers an abstraction of the specific capabilities of the underlying hardware.

\subsection{The GPU Programming Model}
\begin{figure}[ht]
\centerline{\includegraphics[scale=.30]{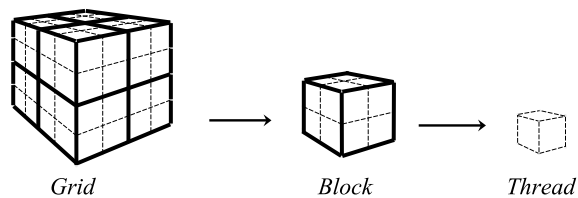}}
\caption{The three-level hierarchy of the GPU Programming.}
\label{fig_constructs}
\end{figure}

The GPU programming model plays an important role in the design and development of GPU accelerated programs, and it is governed by a three-level hierarchy, which defines the parallel work-space of the GPU. This hierarchy corresponds to the thread, block and grid (See Figure \ref{fig_constructs}). 

A thread is an abstract compute element that is in charge of executing the kernel once. A block is a container of threads, and has the property that all of its threads can synchronize among themselves and can share information via cache. Synchronization among blocks is not possible in the current programming model, unless the kernel is terminated and a new kernel is executed, or by introducing newer features such as cooperative groups, but again they are limited in size (grids cannot be as large as normal grids). The grid is the last construct, and it contains all the blocks, spatially identified, that will be employed in the execution of the kernel. Lastly, the kernel is the function that is executed in GPU by all threads. Parallel execution is achieved by executing, simultaneously, several groups of 32 threads. These groups are called \textit{warps} and play an important role in tensor core programming.
With the help of programming tools such as OpenCL or CUDA, one can implement a parallel GPU algorithm that will accelerate data-parallel applications, and this performance will scale automatically in the presence of more capable hardware, \textit{i.e.}, switching a GPU by a more powerful one will lead to automatic performance gains, without the need\footnote{Although automatic scaling occurs, reaching the maximum possible performance of a newer GPU architecture usually requires some small changes in the code.} to change the code.

\subsection{Programming GPU Tensor Cores}
\begin{figure}[ht]
\centering
\includegraphics[scale=.30]{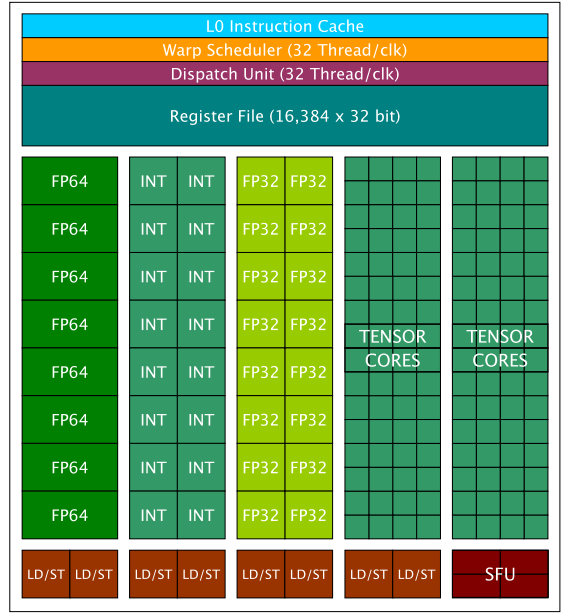}
\caption{A processing group of the Nvidia Tesla V100 which has a total of 640 tensor cores. Image inspired from the Nvidia CUDA programming guide \cite{cuda}.}
\label{fig_gpu_tensor_cores}
\end{figure}

With the latest rise of Machine Learning applications, and more specifically the fast adoption of Deep Learning in multiple fields of science and technology, CPU and GPU Companies have started including application specific integrated circuits (ASICs) to their processors to further accelerate the computational tasks involved in the phases of training and inference in Deep Learning applications \cite{inproceedings,Zhao:2017:ABC:3020078.3021741,5325422,Putnam:2014:RFA:2665671.2665678}. This change has led to the inclusion of \textit{Tensor Core} (TC) units in recent Nvidia GPUs, which are special purpose processing units that sit next to the GPU CUDA cores in the streaming multi-processors (SM) of the chip, as shown in Figure \ref{fig_gpu_tensor_cores}.

The aspect that make tensor cores an attractive feature is the performance it can offer in comparison to the traditional GPU cores. Today, the Nvidia Volta GPU Tesla V100, Quadro V100 and Titan V all include around 640 tensor cores, and they can offer up to 120 TFLOPS in mixed FP16-FP32 precision. In comparison, the traditional CUDA cores, which are 5120 in total for the GPUs recently mentioned, offer up to $\sim 15$ TFLOPS of performance in FP32 precision and around $\sim 7$ TFLOPS in FP64 precision. Indeed, the fast performance of tensor cores is an attractive opportunity to explore possible applications that can take advantage of this new technology. 
For the programmer, the only operation allowed when using tensor cores is the matrix-multiply-accumulate (MMA). Although GPU tensor cores work at hardware level with $4\times 4$ matrices, the programming model exposes the MMA operation in terms of dimensions $m \times n \times k$, \textit{i.e.}, 
\begin{equation}
    D_{m\times k} = A_{m \times n} \times B_{n \times k} + C_{m \times k}
\end{equation}
where $m \times n$, $n \times k$ and $m \times k$ cannot exceed 256 elements. 
An algorithm that takes advantage of tensor core acceleration would redesign its work as multiple MMA operations that occur in parallel.

An important restriction of tensor cores is the numerical precision. Although the $C$ and $D$ matrices can store FP32 numbers, currently 
the operation $A\times B$ is done in FP16 precision. This mixed mode of operation may have negative effects in some applications that have low tolerance to some numerical error. Nevertheless, it is important to note that this extra numerical error can be kept up to a margin considering that all intermediate results can be stored in $C$ and $D$ 
which have FP32 precision, \textit{i.e.}, if the error of $A \times B$ is low, it can remain low for the whole solution of large problems as well. 
One reason of why part of the MMA operation works at FP16 is because in general, Deep Learning applications are not affected by the numerical error introduced by FP16 precision.

GPU Tensor Core programming is currently supported in Nvidia CUDA. A new structure, called \textit{fragment}, is introduced in the programming model. A fragment can be 
considered as a small matrix register that will participate in a MAA operation. Given that four matrices are involved in one MMA, $D = A \times B + C$, 
one would have to declare four fragments in the CUDA code to perform one MMA (at least three if $C = A \times B + C$. Matrix sizes allow some flexibility 
as long as they follow the scheme $D_{m\times k} = A_{m\times n} \times B_{n\times k} + C_{m\times k}$ and the total number of elements for a fragment does not exceed $256$ elements. 
In this work, the size of interest is $m \times n \times k = 16 \times 16 \times 16$. Other configurable parameters for fragments are the data type, column/row major and an explicit parameter to specify if it is matrix $A$, $B$, or an accumulator such as $C, D$. 

Declaration of fragment variables and their operations have a different behavior when compared to the classic GPU programming model. When a fragment is declared in code, all threads of a same warp share this declaration, \textit{i.e.} analogous as how threads from a block share the declaration and scope of a \textit{shared} variable, but this time the scope is at a warp level. When an operation is performed over a fragment, all threads from the corresponding warp cooperate for such operation and synchronize themselves for the next operation. All threads of a warp must meet the fragment operation at the same location in the program, otherwise undefined behavior is expected. Four operations are available to operate fragments, these are the following:
\begin{itemize}
    \item $load\_matrix\_sync$: Loads matrix data from GPU global memory (VRAM) to the matrix fragment $A$ or $B$. Pointer strides and offsets can be given 
    to retrieve the desired 2D region of memory. 
    \item $fill\_fragment$: Fills any matrix fragment given as parameter with a scalar constant number. 
    \item $store\_matrix\_sync$: Stores an accumulator ($C$ or $D$) fragment given as parameter into global memory. Pointer strides and offsets can be given to store at a 
    desired 2D region of memory. 
    \item $wmma\_sync$: Performs the matrix multiply accumulate (MMA) operation. Matrices $A$ and $B$ and at least one accumulator must be given as parameter. 
\end{itemize}
Appendix \ref{app_mma_example} shows an example of a MMA operation for all warps of the grid, with each warp acting on a different region of memory. The Nvidia CUDA Programming Guide provides a more extensive description and explanation of tensor core programming \cite{cuda}.

\section{Related Work}
\label{sec_related_work}
Related works can be classified in two categories: (1) works that study tensor cores or any other ASICs in a general way, and (2) works on parallel reduction. 

\subsection{Works on Tensor Cores and ASICs}
Markidis \textit{et al.} \cite{tensor_cores_2018} studied approaches to program Nvidia Tensor Cores, as well as the performance and the precision loss due to computation in mixed precision. The authors also explain the three levels of programming that can make use of the matrix-multiply-accumulate (MMA) operation: (1) CUDA Warp MMA (WMMA) API, (2) CUTLASS, and (3) cuBLAS GEMM. The tensor core programming is analyzed in different aspects such as programmability, performance and precision. The authors report that the maximum performance obtained was with the cuBLAS GEMM implementation, where they achieved 83 $TFLOPS$ in their test environment (approximately $74\%$ of the theoretical performance), followed by CUTLASS with 62 $Tflops/s$. Their WWMA implementation did not provide any performance improvement, however the authors realized they did not use shared memory in the process, which is an important aspect in order to accomplish efficient tensor core computation. They also observed that when the size of the input matrix increases, the error may increase significantly in some cases. For this, the authors include mechanisms to increase precision such as the Kahan summation and \textit{iterative precision refinement}. Zhe Jia \textit{et al.} have studied the Nvidia Volta and Turing architectures in detail, covering different aspects including the tensor core programability as well \cite{jia2019dissecting, DBLP:journals/corr/abs-1804-06826}.
In more general terms, the Google Tensor Processing Unit (TPU), deployed for data-centers in 2015, is another processor comparable to the tensor cores found in Nvidia GPUs, that accelerates the inference phase of neural networks  \cite{tpu_google_2017}. Norman \textit{et al.} compared the TPU to a server-class Intel Haswell CPU and an Nvidia K80 GPU. They used workloads written with the TensorFlow framework. The results showed that the TPU was on average about 15$\times$ - 30$\times$ faster than its contemporary GPU or CPU, and about 30$\times$ - 80$\times$ higher in TOPS/Watt (Tensor Operations per Second per Watt).

\subsection{Parallel Reduction}
In the context of parallel reductions, CPU-based ones often rely on high-level frameworks or tools such as OpenMP which also offer the reduction pattern \cite{case_study_omp_reduce}. In the case of GPUs, the parallel reduction has been addressed several times. Nickolls, Buck and Garland \cite{nickolls_2008} propose a parallel sum of cost $O(\log_2(n))$ time, where each thread loads one element of the input array, and then adds pairs of values in parallel as $x[i] = x[i] + x[i+p/2]$ where $p$ is the number of threads. The loop in this kernel implicitly builds a summation tree over the input elements where in the end the first data slot of each thread-block holds the reduction result of a sub-set of numbers, the next kernel call would do a reduction on the number of partial results, which turns out to be the number of blocks. 

M. Harris described an efficient CUDA algorithm for the parallel reduction \cite{harris2007optimizing, harris_2005} that was well adapted to the restrictions of the GPU at the time of year 2007. The author illustrates seven different optimizations that are relevant to the parallel reduction, achieving a final version that is up to $30x$ times faster than the initial GPU version presented. Harris mentions that although the time complexity of the parallel reduction is indeed $O(\log_2(n))$, the cost is not efficient if $p=n/2$ threads are used. Instead, one can employ $p = n/{\log_2(n)}$ threads as stated by Brent's Theorem, leading to a parallel efficient cost algorithm. Although the GPU has evolved significantly since that year, the ideas presented are still relevant for many parallel GPU-based algorithms.

Another alternative is to combine a parallel $O(\log_2(n)$ cost algorithm and atomic add operations \cite{cuda}, in such way that each block of threads cooperate for the parallel reduction, and then each block result is combined to form the global result via atomic operations on a unique register. The block-level reduction can also be achieved via warp-level shuffle instructions, which are part of the modern CUDA programming model. In the case of distributed computing, message passing tools such as MPI \cite{omp_ompi} or higher-level frameworks such as MapReduce \cite{mapreduce} allow fast computation of reduction operations. 

In 2018 Carrasco, Vega and Navarro \cite{8705253} proposed a tensor core based parallel algorithm for the reduction problem, and formulated asymptotic upper bounds for its potential running time and speedup. The present work extends the previous one with improvements to the algorithm itself as well as an extensive experimental evaluation; it adds new asymptotic upper bounds for the chained approach, describes and benchmarks three possible variants for implementation with are of high importance on the application side. Furthermore, the manuscript dedicates extensive sections for an in-depth experimental analysis of performance and numerical error under different GPUs and block/chain configurations.

In 2019, Dakkak \textit{et al.} \cite{10.1145/3330345.3331057} developed a reduction and scan algorithm based on tensor core units. For the case of the reduction, they propose an algorithm very similar to the one formulated by Carrasco, Vega and Navarro \cite{8705253}, however some aspects were left out which are considered in this present work, such as (1) a detailed theoretical formulation and analysis of the parallel arithmetic reduction, both chained and un-chained, using tensor cores, obtaining upper bounds that are relevant in the long term (beyond how actual tensor cores work today) such as the non-chained and chained upper bounds. (2) A detailed presentation of performance results under different GPUs, block and chain configurations, which are important when working with tensor cores from different GPU hardware, and (3) experimental results on the numerical precision ($\%$ of error) of the proposed implementations with respect to a CPU reduction using FP64 precision, for both normal and uniform distributions, which are very different scenarios in terms of numerical error accumulation, and represent the type of inputs found in several applications of science and technology. 

The next section presents the new approach from its formulation to its implementation.

\section{Formulation of The Tensor Core Based Arithmetic Reduction}
\label{sec_formulation_analysis}
Given an array of $n$ elements,  $X = \{x_1, x_2, ..., x_n\}$,   
the arithmetic reduction $R(X)$ is defined as
\begin{equation}
    R(X) = \sum_{i=1}^n x_i
\end{equation}
On a serial processor the computation is carried as sequential additions on a single accumulator variable to construct the sum of all elements. 
Such algorithm is considered efficient (as no element can be left out) as it does the necessary and sufficient number operations, and it costs $\Theta(n)$ time steps.
When trying to compute $R(X)$ in parallel, data dependencies and race conditions must be treated within the reduction process. It is known that a parallel reduction 
can sum pairs of values in parallel at each time step, leading to a parallel cost of $T_{p=n/2}(n) = O(\log_2(n))$ using $p=n/2$ processors. At each time step $k$, the size of the 
problem is $\frac{n}{2^{k-1}}$ and $\frac{n}{2^k}$ threads are sufficient to do the step in parallel taking pairs again. In general, \textit{i.e}, the $i$-th thread 
sums the elements $x_i$ and $x_{i + n/{2^k}}$ and stores the partial sum in $X[i]$. Such parallel step costs $O(1)$ time and cuts the problem in half. 
Consecutive applications of this process lead to the recurrence
\begin{align}
T(n) &= O(1) + T(n/2)
\end{align}
subject to $T(2) = O(1)$, 
which by the master theorem it can be shown that it is $T_{p=n/2}(n) = O(\log_2(n))$. When considering the parallel cost, which is defined as $C_p = T_p(n) \cdot p$, with $p$ the number of processors (assuming 1 thread running on one 1 processor) employed, then using $n/2$ processors leads to a parallel cost of
\begin{equation}
C_p(n) = \log_2(n) \frac{n}{2} = O(n \log_2(n))
\end{equation}
which makes the algorithm cost inefficient as it is greater than the $O(n)$ cost of the sequential algorithm. To improve on this cost, one can use Brent's Theorem \cite{Brent_1974} which states the inequality
\begin{equation}
T_p(n) \le \frac{T_1(n)}{p} + T_{\infty}(n)
\end{equation}
where in the case of the arithmetic reduction we have that $T_1(n) = O(n)$ and $T_{\infty}(n) = \log_2(n)$. With this, one can choose the number of processors as $p = \frac{n}{\log_2(n)}$ without sacrificing parallel time, \textit{i.e}, 
\begin{equation}
    T_p(n) \le 2\log_2(n) = O(\log_2(n))
\end{equation}
With this change, the cost of the parallel algorithm is now efficient, \textit{i.e},
\begin{equation}
    C_p(n) = \log_2(n) \frac{n}{\log_2(n)} = O(n)
\end{equation}

It is important to consider that there is an asymptotic lower bound of $\Omega(\log_2(n))$ for the parallel reduction. Nonetheless, one can 
still improve in the constants involved as in the base of the logarithm, which can make an important difference in experimental performance. Such is the case of reducing with tensor cores, where a large number of arithmetic operations can be encoded into MMA operations and executed in just one GPU cycle, which is equivalent to one time unit. The following sub-section presents the new algorithm for parallel arithmetic reductions using tensor cores operations and analyzes its cost as well as speedup over the parallel reduction algorithm described recently. 

\subsection{Encoding Additions Into Tensor Core MMAs}
The tensor core programming model exposes a single operation to the programmer, the matrix-multiply-accumulate (MMA). That is, given three matrices $A, B, C$, the MMA operation computes
\begin{equation}
    D = A \times B + C
\end{equation}
In one GPU cycle. The tensor core computing model allows many MMA operations to occur simultaneously in parallel. It is interesting to note that in the programming model the tensor core MMA operation is exposed in terms of $m \times n \times k$ and allows the definition of matrices of size $16 \times 16$ to the programmer, even when the actual operation at hardware level is carried in terms of $4 \times 4$ matrices. The process of splitting the $16 \times 16$ workload into smaller $4 \times 4$ works is done automatically by the GPU scheduler, but splitting a large problem of size $n$ into several $16 \times 16$ matrices is not automatic and must be designed manually. This last aspect is the one important for the research, as it is related to the research question of whether a reduction problem of size $n$ can be encoded into multiple MMA operations. The presentation of the new reduction algorithm will proceed in terms of $m \times m$ MMA matrices, as it favors the analysis in the next sub-section. 

The intuition behind a tensor core MMA based reduction is to produce many partial summations of groups of $m^2$ numbers, in parallel. To achieve this, we employ two MMA operations on each group of $m^2$ values, but later on this section the idea extends to an $R$-chain of MMA operations. For the first MMA operation, defined as $D = A \times B + C$, $m^2$ elements of the input array $X[\ ]$ are loaded in $B_{m\times m}$ in parallel such that $B_{m,m}$ is actually the $m^2$-th element of the group. Other warps of the grid will access their corresponding region of global memory by applying an offset of $B_x|B| + |w|*m^2$ where $B_x$ is the block index, $|B|$ is the block-size and $|w|$ the warpsize. Matrix $A_{m \times m}$ is set as an all-ones matrix, also in parallel, and $C$ is a zero-matrix, also set in parallel. 

In order to allow a posterior theoretical analysis of the algorithm, it is assumed that the loading process as well as the eventual MMA operation take place in all warps of all blocks at the same time in parallel, regardless of the fact that in practice there is an internal scheduler to distribute the workload into the finite number of GPU tensor cores. Such assumption is common when studying an algorithm under a PRAM-like computing model, which is the case for this work.
When the MMA operation is executed on $A, B$ and $C$, the result is a matrix, namely $D_{m \times m} = A_{m \times m} \times B_{m \times m} + C_{m\times m}$, of the form 
\begin{align} \label{eq_MMA1}
D &= 
\begin{bmatrix}
  \mbox{\huge1}\\
\end{bmatrix}_{m \times m}
\times
\begin{bmatrix}
    x_{11} & \dots  & x_{1m} \\
    \vdots & \ddots & \vdots \\
    x_{m1} & \dots  & x_{mm}
\end{bmatrix}
+
\begin{bmatrix}
  \mbox{\huge0}\\
\end{bmatrix}_{m \times m}
\\
&=
\begin{bmatrix}
    \sum_{i=1}^{m} x_{1i}  & \dots &  \sum_{i=1}^{m} x_{1i}  \\
    \vdots & \ddots & \vdots \\
    \sum_{i=1}^{m} x_{mi}  & \dots &  \sum_{i=1}^{m} x_{mi}  
\end{bmatrix}
\end{align}
where each column $D_{k,j}, k = 1..m$ holds the whole set of partial summations from the group of $m^2$ elements.

The second MMA operation is in charge of reducing the partial summations found in the columns of $D$, now into a single value. It is relevant to notice that dealing with one column of $D$ is sufficient as it holds all the partial summations. The other columns of $D$ hold copies of the result. At the same time, given the rigidness of the MMA operation, it takes less time to process the whole matrix within the MMA operation instead of filtering or doing extra considerations to just process a single column. As long as the result of one column is not compromised, doing a full MMA operation is still efficient as it preserves the one GPU cycle cost per MMA. 

The second MMA operation proceeds by changing the order of the multiplying matrices, and re-uses the output matrix $D$ in the position of $A$, while using $A$ in the position of $B$. With these changes, the second MMA operation becomes the following expression 
\begin{align} \label{eq_MMA2}
D' &= 
\begin{bmatrix}
    \sum_{i=1}^{m} x_{1i}  & \dots &  \sum_{i=1}^{m} x_{1i}  \\
    \vdots & \ddots & \vdots \\
    \sum_{i=1}^{m} x_{mi}  & \dots &  \sum_{i=1}^{m} x_{mi}  
\end{bmatrix}
\times
\begin{bmatrix}
  \mbox{\huge1}\\
\end{bmatrix}
+
\begin{bmatrix}
  \mbox{\huge0}\\
\end{bmatrix}
\\
&=
    \begin{bmatrix}
    \sum_{i=1}^{m}\sum_{j=1}^{m} x_{ij}   &  \dots  & \sum_{i=1}^{m}\sum_{j=1}^{m} x_{ij}  \\
    \vdots & \ddots & \vdots \\    
    \sum_{i=1}^{m}\sum_{j=1}^{m} x_{ij}   &  \dots  & \sum_{i=1}^{m}\sum_{j=1}^{m} x_{ij}  \\
\end{bmatrix}
\end{align}
The resulting matrix $D'$ contains the reduction of the $m^2$ numbers, replicated in all of its elements. Once the second MMA operation finishes, the thread in charge proceeds to write the reduction result, \textit{e.g}, $D'_{1,1}$, into global memory as part of the new array of partial sums.

The global idea of the algorithm is to subdivide the domain of $n$ numbers into $\frac{n}{m^2}$ blocks of $m^2$ elements and execute the 2-step MMA reduction proposed for each group in parallel. This reduces the problem size by a factor of ${m^2}$. In the next iteration, the idea is to take the smaller version of the problem, of size $n' = \frac{n}{m^2}$ and reduce it again with the 2-step MMA operations for all the possible groups of $m^2$ elements. This process is carried iteratively until the reduction set fits in just one group of $m^2$ elements, for which a final tensor core reduction returns the result of the whole reduction problem of size $n$.

This new tensor core MMA based reduction, namely $R_{tc}(X)$, with $X = \{x_1, \dots, x_n\}$, can be described by the following recurrence
\begin{align}
\label{eq_reduction_recursive}
R_{tc}(X) &= R_{tc}(M(x_1..x_{m^2}),\dots, M(x_{(k-1)m^2+1}..x_{km^2}))
\end{align}
where $M(...)$ is the tensor core based MMA reduction and the initial condition 
is defined as
\begin{equation}
    R_{tc}(x_1..x_{m^2}) = M(x_1..x_{m^2})
\end{equation}
The process described is summarized in Algorithm \ref{alg_mma}, and the kernel is summarized in Algorithm \ref{alg_kernel}.
\begin{algorithm}
\SetAlgoLined
\KwResult{sum of $n$ numbers}
X = \{$x_1, x_2, ..., x_n$\}\;
n = $|X|$\;
\While{$n \ge m^2$}{
  KernelMMA(X, n)\; 
  n := $\lceil \frac{n}{m^2} \rceil$\;
}
\caption{Tensor Core Based Reduction}
\label{alg_mma}
\end{algorithm}

\begin{algorithm}
\SetAlgoLined
\KwData{X, n}
fragment A, B, C\;
$offset = m^2 \cdot (\frac{t_{id}}{|warp|} + Block_{id} \cdot \frac{|Block|}{|warp|})$\;
\If{$offset < n$}{
    setFragment(A, 1.0)\;
    setFragment(C, 0.0)\;
    loadMatrix(B, X$+offset$)\;
    MMA(C, A, B, C)\;
    copyFromTo(C, A)\;
    setFrament(B, 1.0)\;
    setFrament(C, 0.0)\;
    
    MMA(C, A, B, C)\;
    $X[\frac{offset}{m^2}] = C_{0,0}$\;
}
\caption{KernelMMA}
\label{alg_kernel}
\end{algorithm}
An eventual implementation of this algorithm would have to manage border conditions such as when $n$ is not a power of $m^2$ and 
perform register-level operations when moving fragment elements from $C$ to $A$. The following sub-section presents a fine analysis 
(considering constants) of the proposed reduction algorithm.

\subsection{Analysis of Running Time and Speedup}
A simplified GPU Computing model is used to analyze the running time and speedup of the algorithm with respect to a classic parallel reduction.
This computing model is based on the PRAM (Parallel processors and one large shared memory) with extra 
considerations related to the GPU architecture. Under this model, the costs for the different types of parallel operations are:
\begin{itemize}
    \item Coalesced r/w operations on global memory cost $1$ unit of time.
    \item non-Coalesced r/w operations on global memory cost $w$ units of time. 
    \item The cost of r/w operations on registers or L1 cache is at least an order of magnitude less than r/w operations on global memory. 
    \item Parallel tensor core MMA operations cost $1$ GPU cycle, therefore 1 unit of time.
    \item Transfers between global memory and tensor core fragments cost $1$ unit of time.
\end{itemize}
In the case of a GPU reduction, it is possible to load a continuous region of $m^2$ elements into the tensor core fragments, which reside in cache memory near the tensor cores. This loading of information takes a unit of time. Then, the algorithm proceeds and executes the 2-step tensor core MMA reduction simultaneously for all the $m^2$ groups that can be made for the array $X[1..n]$. In between these two MMA operations, internal register/L1 Cache operations are carried to move data from fragment C to A, which account for 1 extra unit of time as they are much faster than global memory operations. Lastly, once the reduction of a group is done, the thread in charge writes this result in its corresponding location in $X$, taking another unit of time. Combining these costs into one expression leads to the following recurrence 
\begin{equation}
    T_{tc}(n) = 5 + T_{tc}(\frac{n}{m^2})
\end{equation}
that stops at $T_{tc}(m^2) = 5$. Solving the recurrence leads to the final cost of
\begin{equation}
    T_{tc}(n) = 5\log_{m^2}(n)
\end{equation}
A standard parallel GPU-based reduction, under the same cost model, would take one unit of time for reading the first element of each thread (all in parallel),  another unit of time for reading the second element, a unit of time for adding two numbers, and another unit of time for storing the value back to global memory. The recurrence for such algorithm would be $T(n) = 4 + T(n/2)$ which is equal to $T(n) = 4\log_2(n)$. When comparing the two running times, the potential speedup for the tensor core based reduction becomes
\begin{equation}
    S = \frac{T(n)}{T_{tc}(n)} = \frac{4\log_2(n)}{5\log_{m^2}(n)} = \frac{4}{5}\log_2(m^2).
\end{equation}
A value of $m \ge 2$, which is the minimum value $m$ could take in order for the reduction to work, is already sufficient to provide an acceleration of $S > 1$. 
In actual GPU hardware, the value is $m=4$, which would lead to an even greater speedup. This Section has proven that under a PRAM-like model, the tensor core based reduction 
will indeed provide a speedup over the standard GPU reduction algorithm.

\subsection{Chained MMA Operations per Warp}
It is possible to generalize the proposed algorithm and obtain a more general theoretical running time. The intuition behind is to relax the idea of having to perform two MMA operations 
to reduce $m^2$ numbers and instead generate a sequential chain of $R$ MMA operations per warp. Such approach can be expressed as
\begin{align}
    C_1 &= \begin{bmatrix}
            \mbox{\huge1}\\
          \end{bmatrix} \times M_1 + C_0\\
    C_2 &= \begin{bmatrix}
            \mbox{\huge1}\\
          \end{bmatrix} \times M_2 + C_1\\
          \dots\\
    C_R &= \begin{bmatrix}
            \mbox{\huge1}\\
          \end{bmatrix} \times M_R + C_{R-1}\\
    C_{R+1} &= C_R \times 
          \begin{bmatrix}
            \mbox{\huge1}\\
          \end{bmatrix} + 
          \begin{bmatrix}
            \mbox{\huge0}\\
          \end{bmatrix}
\end{align}
where $M_1, M_2, ..., M_R$ are consecutive groups of $m^2$ numbers in global memory, \textit{i.e.}, $X = \{M_1,M_2,..., M_R, M_{R+1}, ..., M_{n}\}$ with $R \ge 1$ and $n >> R$. 
The last $(R+1)$-th MMA operation is analog to the second MMA in the two-MMA design presented before, which is required for the summation to be reduced to a single value.
An advantage of executing a chain of $R$ MMAs per warp is that the total number of MMA operations is practically half than in the previous version of the proposed algorithm, 
where two MMAs were performed per group of $m^2$  elements, \textit{i.e.}, only $R+1$ MMA operations are required for reducing $Rm^2$ numbers, instead of $2R$ as before. However, 
these operations are sequential for the warp, therefore there exists a tradeoff when the number of available processors is finite. 
Another advantage of the chained approach is that the $R$ chains do not need to perform register/L1-cache level operations as these are only required before the last MMA. Again, 
this improvement may only be seen with a finite number of processors. The total cost of algorithm using the chained MMA improvement, under the Simplified GPU Computing model, 
can be expressed as the recurrence
\begin{equation}
    T_{tc}^R(n) = (2R+1 + 2) + T_{tc}^R(\frac{n}{Rm^2})
\end{equation}
where the term $2R$ corresponds to the load and MMA operation involved in each of the $R$ steps, the $1$ value corresponds to the final MMA, and the $2$ is the cost 
of internal register/cache operations and the final store on global memory. The recurrence is equal to
\begin{equation}
    T_{tc}^R(n) = (2R+3)\log_{Rm^2}(n).
\end{equation}
If $R=1$, the result is the same as the initial version of the algorithm, and when $R>1$, the running time shows a higher value, however this model assumes an infinite number of processors are available. In order to find an optimal value of $R$, experimental evaluation is required. The next Section presents the performance of different variations of the algorithm, and finds optimal values for $R$.

\section{Implementation and Variants}
\label{sec_variants}
Three different variations of the proposed algorithm were implemented and tested. The implementation is available at \url{https://github.com/crinavar/xxxxx}\footnote{The download link will be made available to the community in the published version of the article.}.

\subsection{Variant \#1: Recurrence}
\begin{figure}[ht!]
    \centerline{\includegraphics[scale=.7]{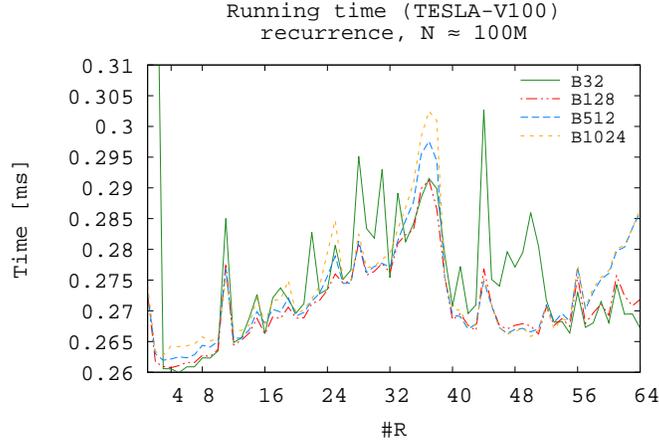}}
    \caption{Performance of the recurrence variant under different block sizes and values of $R$.}
    \label{fig_recurrence-rconf}
\end{figure}

This variant is a direct CUDA bottom up implementation of the two-step MMA recursive method described in Algorithms \ref{alg_mma} and \ref{alg_kernel}, and expressed in Eq. (\ref{eq_reduction_recursive}), with the inclusion of the chained MMAs feature.
Different configurations of $B,R$ were tested in order to find the values that lead to the best performance. Figure \ref{fig_recurrence-rconf} shows the performance of different $B,R$ configurations for an input of $\sim 1$ million numbers following a normal distribution $[\mu = 0, \sigma^2 = 1]$. 

From the plot one can observe that the curves follow an irregular sawtooth shape, and the fastest performance is obtained with $B=32, R=5$ followed by the configuration $B=128,R=4$. Future mentions of this variant use the $B=32,R=5$ setting. 
It is worth mentioning that this variant was initially implemented in two ways; (1) iterative kernel calls within a CPU loop and (2) CUDA cooperative groups to handle the iterations inside a single kernel launch. The first version was kept as it performed faster than the second. In addition, the second version had limitations on the grid size, because of the cooperative groups feature, which directly limited the range of the problem size.

\subsection{Variant \#2: Single Pass}
\begin{figure*}[ht!]
    \centerline{\includegraphics[scale=.45]{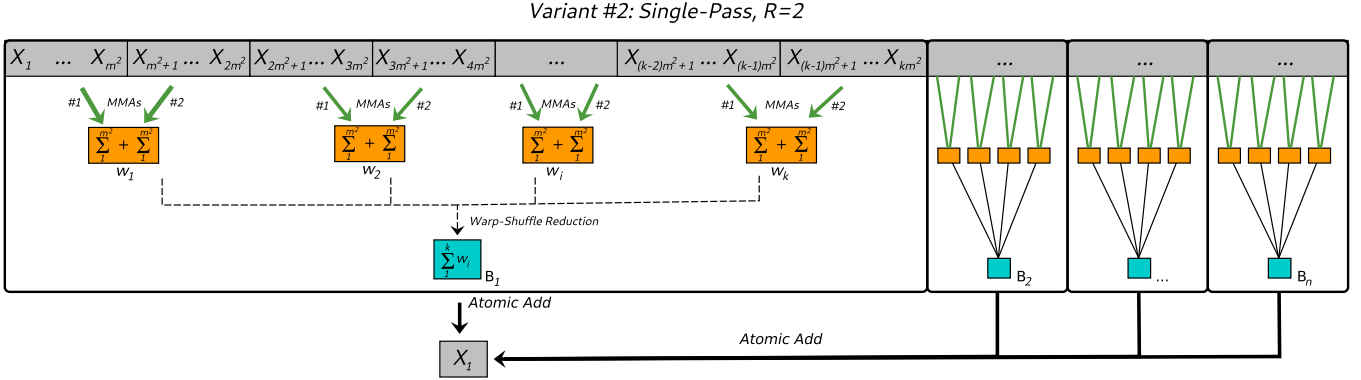}}
    \caption{Illustration of the Chained MMAs variant. The grey regions are global memory locations, while the orange and blue ones correspond to on-chip memory. The reduction starts with each warp of each block performing a sequence of $R=2$ accumulated MMA operations. Then, a single value per block is computed through a single warp-shuffle reduction of at most 32 elements. Lastly, all block results are combined via atomic additions on the first memory location of the global memory. }
    \label{fig_reduction_block}
\end{figure*}

This variant implements the reduction as a single-pass kernel that combines three levels of computation: (1) chained MMAs, (2) warp-shuffle reductions and (3) atomic operations.
First, $R$ parallel sequences of MMA operations occur for each warp in each block. Then, each block, using its first warp, reduces the warp results into a single value with one warp-shuffle reduction process. Lastly, all blocks reduce their results, using the first thread on each block, through atomic operations on the first global memory location. These three stages occur in a single kernel pass. 
Figure \ref{fig_reduction_block} illustrates the Chained MMAs approach, with its three types of computations involved in the reduction.

\begin{figure}[ht!]
    \centerline{\includegraphics[scale=.7]{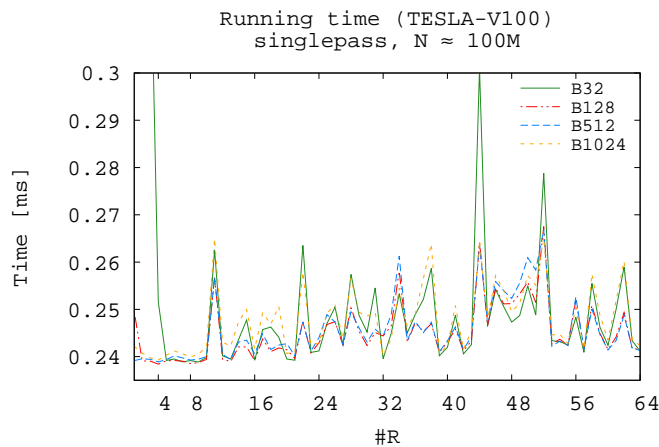}}
    \caption{Performance of the single pass approach under different block sizes and values of $R$.}
    \label{fig_single-pass-rconf}
\end{figure}

In principle, the atomic operations do not incur into a considerable performance penalty because the number of block results is significantly smaller than the original problem size. This effect is further increased for larger values of $R$ in the MMA chains.
As with Variant \#1, different configurations of $B,R$ were tested as well. Figure \ref{fig_single-pass-rconf} shows the performance of different $B,R$ configurations for an input of $\sim 1$ million numbers following a normal distribution $[\mu = 0, \sigma^2 = 1]$. From the plot one can observe again an irregular sawtooth shape that is more pronounced for larger blocks, and less for smaller ones. The fastest configuration for this variant is obtained when using the configuration of $B=128, R = 4$. Future mentions of this variant keep these values. 

\subsection{Variant \#3: Split in Two}
\begin{figure}[ht!]
    \centerline{
    \includegraphics[scale=.7]{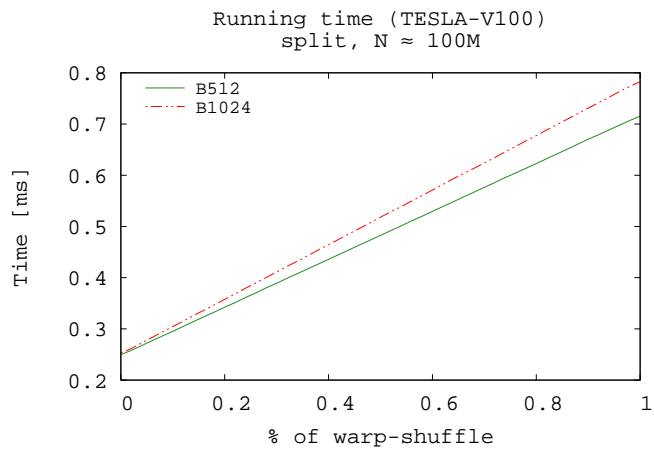}
    }
    \caption{Performance of the split variant under different fractions of warp-shuffle reduction.}
    \label{fig-split-pconf}
\end{figure}

The third and last variant works under the assumption that while tensor cores are being used, classic GPU cores are potentially available to be used at the same time, providing more computational resources per second. Under this assumption, the implementation splits the data domain into two halves, the first one, being a fraction of $f$ from the whole domain, is for tensor core reduction using a chain of $R=1$, while the second half, with a fraction of $1-f$, is for the classic GPU reduction using warp shuffle instructions. Similar to variant $\#2$, the whole reduction is done within a single kernel pass. Figure \ref{fig-split-pconf} shows the performance of the split variant for different fractions of warp-shuffle reductions.

\subsection{Variants Performance and Error}
\begin{figure*}[ht!]
      \centering
      \begin{subfigure}[b]{0.49\textwidth}
      \includegraphics[scale=.65]{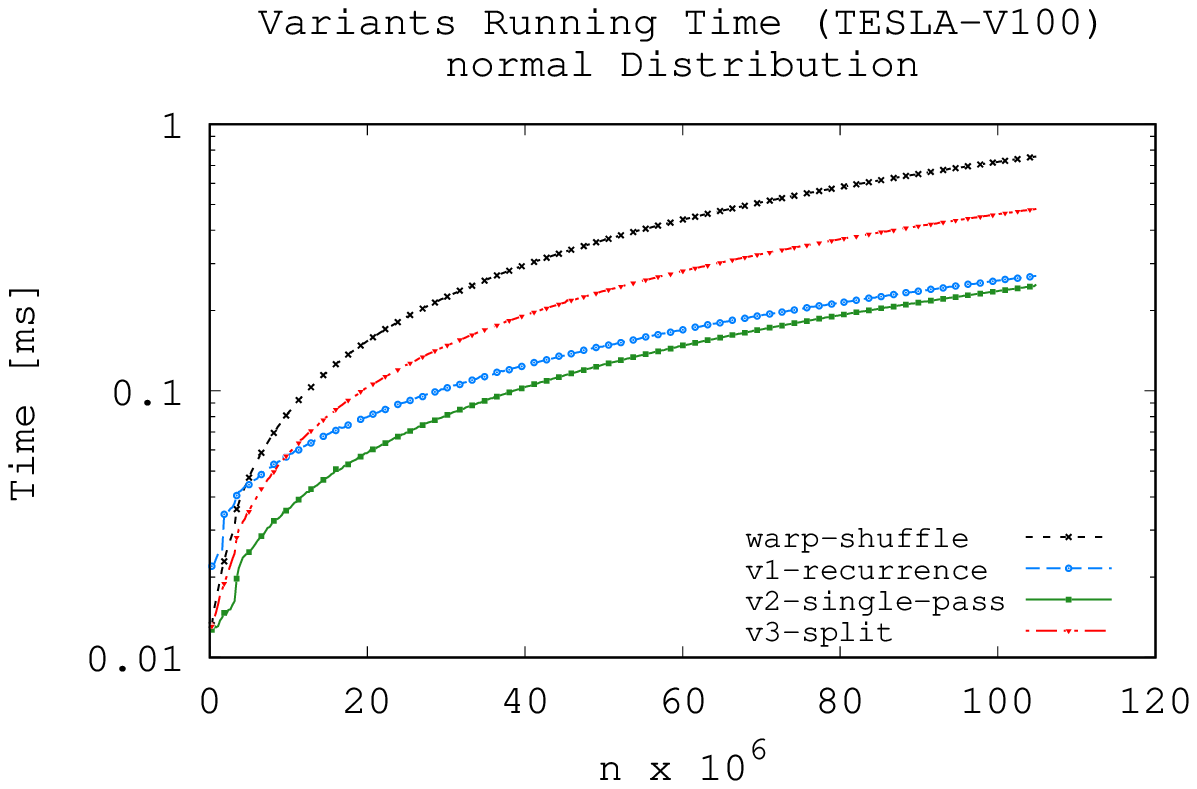}
      \end{subfigure}
      \begin{subfigure}[b]{0.49\textwidth}
      \includegraphics[scale=.65]{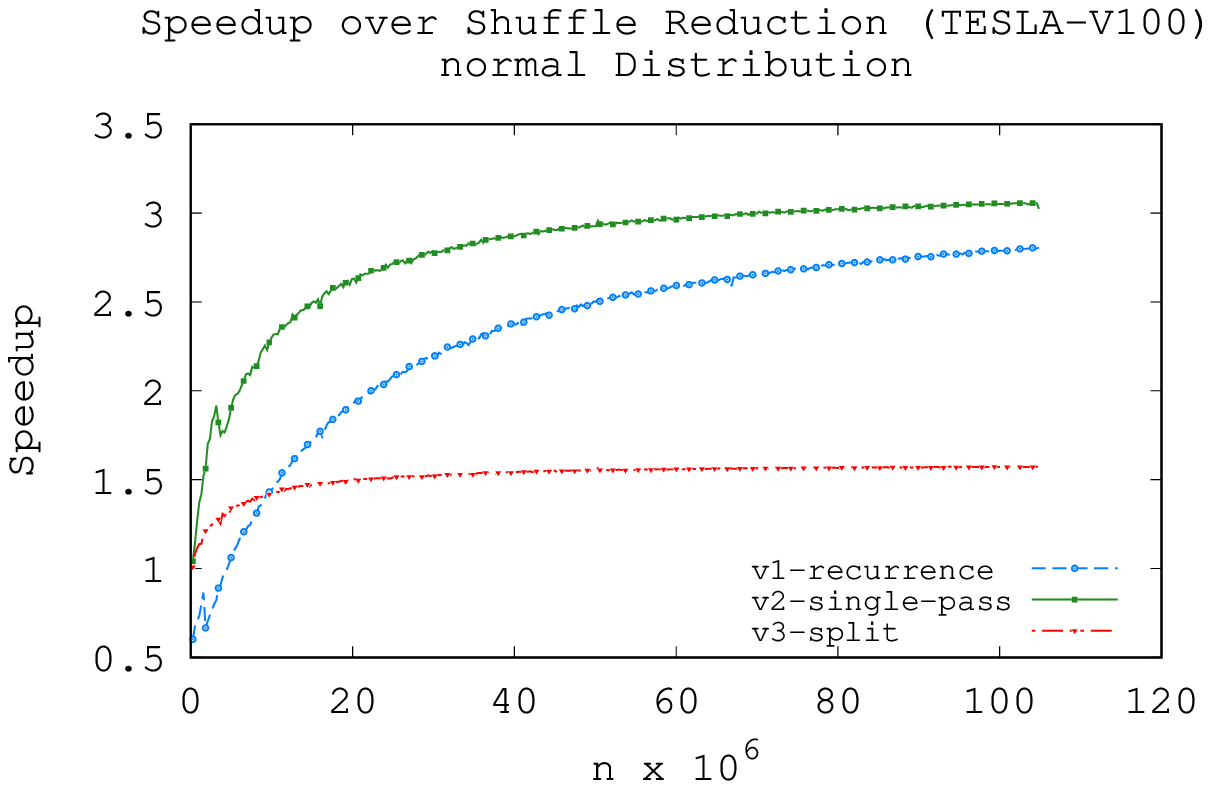}
      \end{subfigure}
      \\
      \begin{subfigure}[b]{0.49\textwidth}
      \includegraphics[scale=.65]{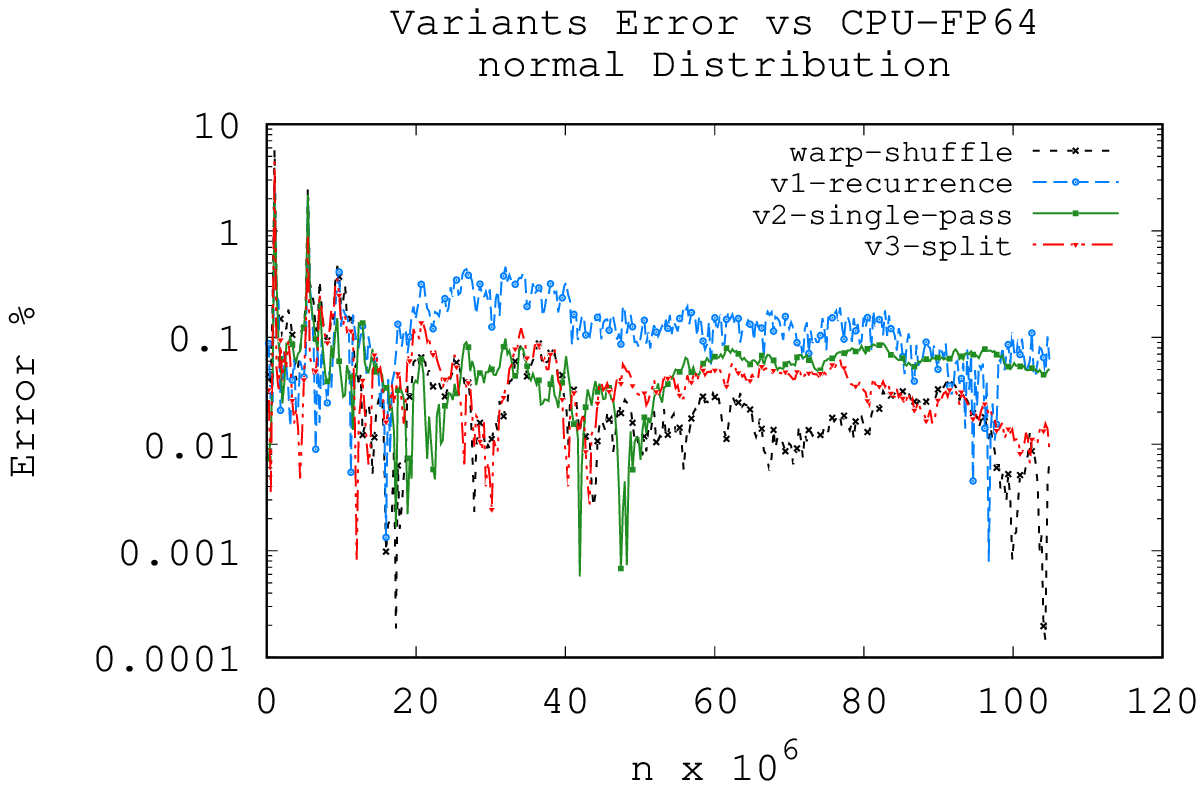}
      \end{subfigure}
      \begin{subfigure}[b]{0.49\textwidth}
      \includegraphics[scale=.65]{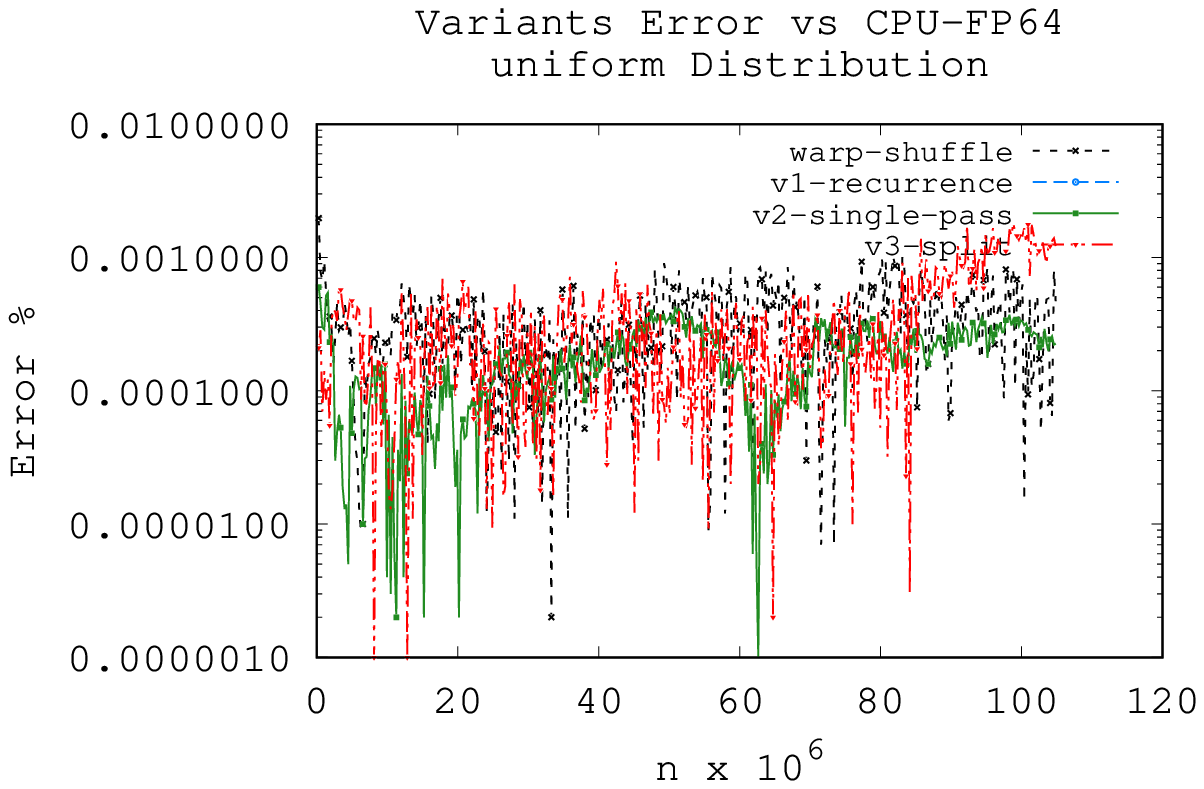}
      \end{subfigure}
      \caption{Runtime, speedup and numerical error of the single-pass variant and CUB reductions.}
      \label{fig_variants_speedup_error}
\end{figure*}

The three variants are compared regarding two aspects: (1) speedup over a classic warp-shuffle reduction (does not use tensor cores, just regular CUDA FP32 cores) and (2) numerical error with respect to a CPU reduction using double precision. The tests were run on a TESLA V100 GPU, and additional performance results using a TITAN RTX can be found in Appendix \ref{app_results_rtx}. \textit{Note: the fastest variant found in this subsection is then compared with Nvidia's CUB library in Section \ref{sec-comparison-cub}}.
Figure \ref{fig_variants_speedup_error} shows the speedup of all variants with respect to a warp-shuffle reduction as well as their numerical error with respect to a CPU reduction in FP64 mode.
From the runtime curves one can note that the single-pass is the fastest variant, closely followed by the recurrence variant, which is slow for small $n$, but catches up in performance for large $n$. On the other hand, the split approach is fast for small $n$ but then becomes the slowest variant for large $n$. In terms of speedup (top-right), the small gap between the single-pass and recurrence variants is more clear, with the single-pass achieving close to $4.5\times$ of speedup, while the recurrence speedup is only slightly above $4.0x\times$ is speedup. The split variant achieves almost $1.5\times$ of speedup over the warp-shuffle approach. The split variant, although promising in theory, did not perform as fast as expected. It is probable that the assumption of the GPU being able to use its two types of processors (INT32/FP32 cores and Tensor cores) simultaneously may not be true, or other parameters not considered introduce a penalty in performance.

In terms of numerical error, in the normal distribution $[\mu = 0, \sigma^2 = 1]$ test (bottom left) all variants present less than $1\%$ of numerical error with respect to the CPU reduction, once the input size is $n \ge 10 \times 10^6$ numbers. These errors are in an order of magnitude close to the warp-shuffle reduction error which uses FP32 computation through all of its stages. For the uniform distribution $[0,1]$ the variants present less numerical error, except for the recurrence variant which reached overflow for early values of $n$, because it does not store partial results in FP32 precision.

Although in principle both the single-pass and recurrence variants are promising in terms of performance, the recurrence variant can only work in restricted numerical scenarios, such as a normal distribution, but not in any arbitrary one where the final result tends to increase (or decrease) as $n$ grows. On the other hand, the single pass variant combines sub-results in FP32 precision, allowing it to be used in arbitrary scenarios without generating early overflow or underflow. For these reasons, the chosen variant for further comparisons is the single-pass variant.

\section{Comparison with NVIDIA's CUB library}
\label{sec-comparison-cub}
The fastest variant found in the previous Section, \textit{i.e.}, the single-pass, is compared in terms of performance and error with respect to Nvidia's CUB library \cite{CUB} in both FP16 and FP32 precision modes. CUB is is a software tool from Nvidia that offers several parallel computing patterns optimized for fast performance. One of the functions provided is the arithmetic reduction.

\begin{figure*}[ht!]
      \centering
      \begin{subfigure}[b]{0.49\textwidth}
      \includegraphics[scale=.70]{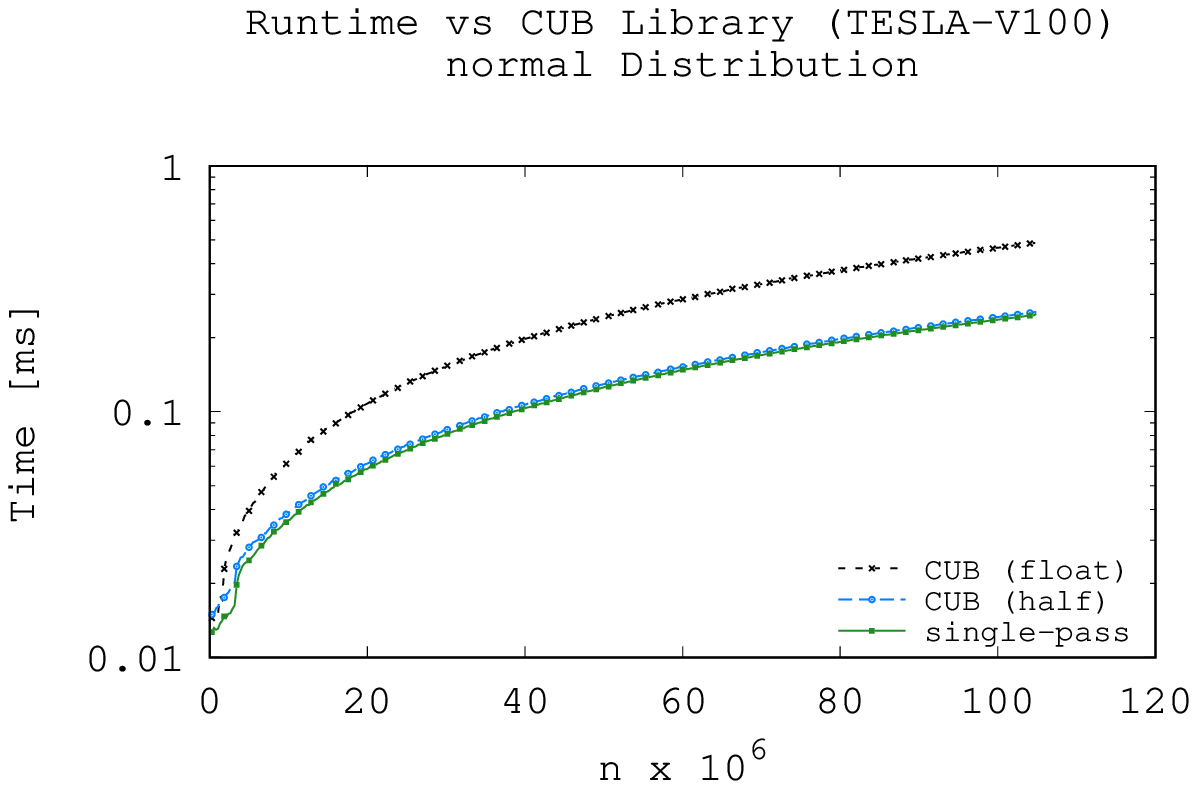}
      \end{subfigure}
      \begin{subfigure}[b]{0.49\textwidth}
      \includegraphics[scale=.70]{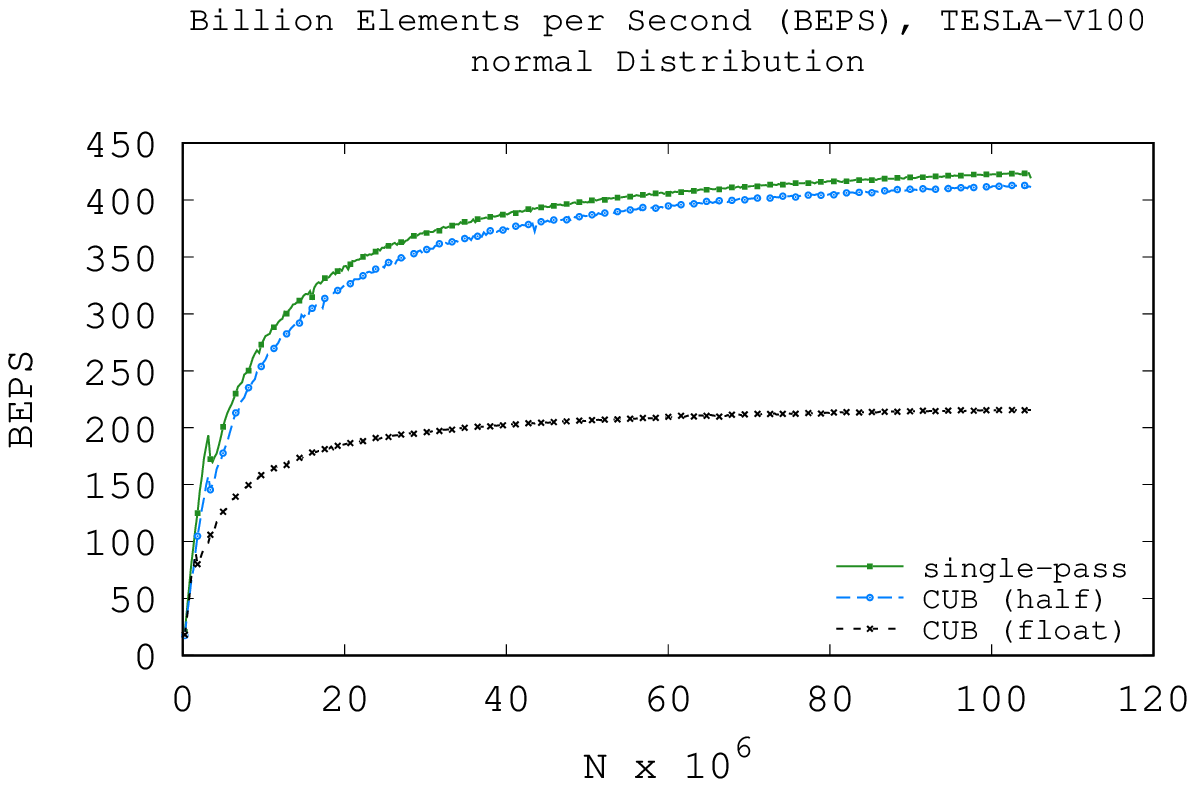}
      \end{subfigure}
      \\
      \begin{subfigure}[b]{0.49\textwidth}
      \includegraphics[scale=.70]{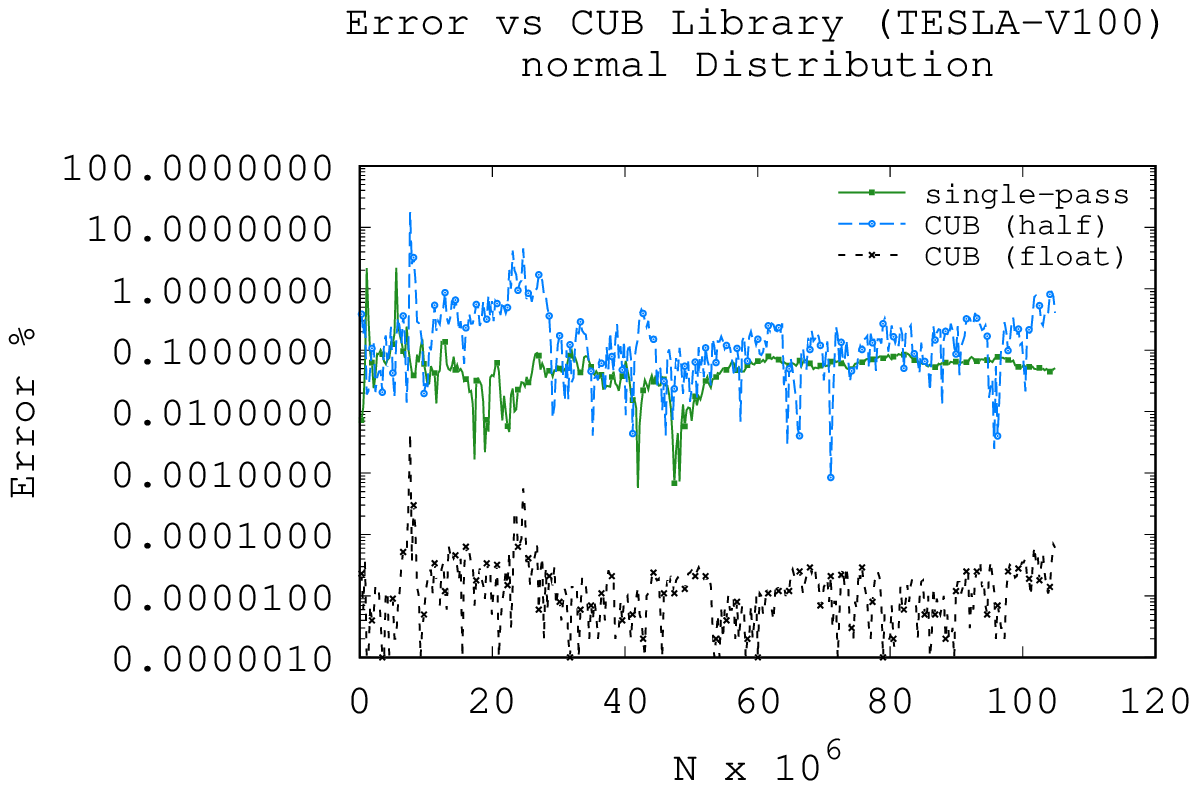}
      \end{subfigure}
      \begin{subfigure}[b]{0.49\textwidth}
      \includegraphics[scale=.70]{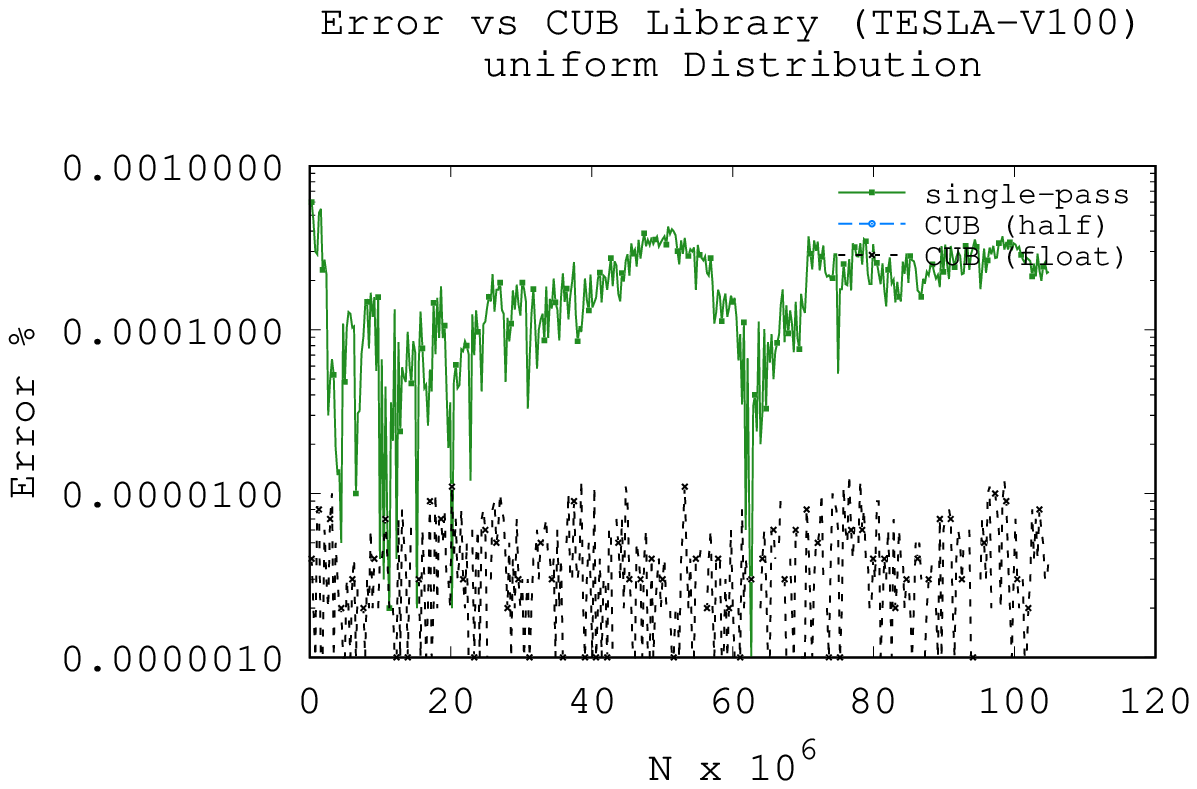}
      \end{subfigure}
      \caption{Runtime, BEPS and numerical error (normal and uniform) of all variants with respect to a warp-shuffle reduction.}
      \label{fig-comparison}
\end{figure*}

Figure \ref{fig-comparison} includes (1) runtime, (2) billion of elements per second (BEPS), (3) the percentage of numerical error with respect to a double-precision reduction on CPU, both in normal and uniform input distributions. The GPU used for these tests is a TESLA V100. Additional performance results using a TITAN RTX are included in Figure \ref{fig_app_comparison_rtx} from Appendix \ref{app_results_rtx}.
From the runtime curves one can notice that the single-pass variant is the fastest of three reduction methods, closely followed by CUB in FP16. In terms of BEPS, the single-pass achieves $\sim 420$ BEPS. The half and float versions of CUB achieve $\sim 400$ and $\sim 200$ BEPS, respectively. In terms of error percentage, for the normal distribution, the float computation of CUB-float is some orders of magnitude more precise than the single-pass and CUB-half, nevertheless these last two stabilize to $\sim 0.1\%$ as $n$ becomes larger. For the uniform distribution, the single-pass and CUB-float have both an error below $0.001\%$. The CUB-half version did not pass this test as the result overflowed at an early stage in the computation.

\section{Conclusion and Discussion}
\label{sec_discussion}
The main contribution of this work is the presentation of a GPU-based arithmetic reduction that exploits the high performance of GPU tensor cores, from its formulation and analysis, to its implementation and experimental evaluation. From the analysis, the main results are the asymptotic time of the tensor core based algorithm, which is $T_{tc}(n) = 5\log_{m^2}(n)$, and the speedup expression with respect to a classic parallel GPU reduction, which is $S = (4/5)\log_2(m^2)$. It is interesting to note that the minimum value that makes the tensor core based reduction work, \textit{i.e.} $m=2$, already produces a speedup of $S > 1$ which is an improvement. Current GPUs such as the TESLA V100 perform $4\times 4 \times 4$ MMA operations in one cycle at hardware level, meaning that in the present we have $m=4$. 
An important result in this work is the fact that the experimental speedup achieved in practice by the single-pass variant practically matches the theoretical speedup when considering $m=4$ which is $S \approx 3.2$. Somehow, the GPU computing cost model used in the analysis, while being simple, can still provide useful insights on what experimental performance to expect for an eventual implementation of a tensor-core based algorithm.

The chained MMA design is an important feature for all three variations proposed, as the best performance for the recurrence variant was $R=5$ with Block size $B=32$ and for the single-pass it was $R=4$ with $B=128$. In both cases, the chain value was $R \ge 1$ which could not have been found without considering a chain of MAAs. The tuning for best block size is also an important aspect to include in this work because we found that different GPU architecture work best using different block sizes (see Appendix \ref{app_results_rtx}).
From the three variations proposed, the single-pass became the fastest and most convenient one in terms of numerical error with respect to a double precision CPU reduction. When compared to a classic warp-shuffle reduction, the single-pass variant achieved over $3\times$ of speedup, and when compared with CUB, it performed slightly faster than CUB-half and $2\times$ faster than CUB-float, while not suffering the overflow effects of CUB-half. It is also important to note that when running the same tests with a TITAN RTX GPU (see Appendix \ref{app_results_rtx}), the speedup was greater.

Although the recurrence variant was discarded because it did overflow with inputs following a uniform distribution, such as numbers uniformly distributed in the range $[0, 1]$, it is still important to note that this implementation may perform as fast or even faster than the single-pass approach for very large values of $n$, as this is often the case of asymptotic logarithmic behaviors, \textit{i.e.}, they may require very large problem sizes for the asymptotic behavior to manifest. Although it is incapable of handling uniform distributions, the recurrence variant may still be useful for reductions that do not generate overflow, such as inputs that follow a normal distribution.

The split approach seemed to be a promising variant in theory as it considered the use of regular CUDA cores with tensor cores at the same time. However in practice this variant was the slowest of the three proposed. It seems relevant to consider a deeper study to understand better in what conditions different kinds of GPU cores can work simultaneously, specially when considering the actual trend in technology to include ASICs of different purposes, such as the tensor cores and recently ray tracing cores.

The results obtained in this work have shown that there is a potential performance that can be exploited from GPU tensor cores, with applications in fields that are not necessarily related to machine learning such as simple reductions, which are a fundamental tool for computational physics (n-body simulations, spin lattice models, PDEs). The three key aspects for finding a fast reduction based on MMAs were (1) the formulation, (2) the asymptotic analysis and (3) the profiling to explore the optimal values of $R$ and $B$. In the future, we are interested in studying more computational patterns that can take advantage of tensor cores or ray tracing cores, such as accelerate GPU thread mapping or simulations in structured/unstructured spaces, as well as to perform a deeper study to understand how tensor core units can work simultaneously with regular cores. 

\section*{Acknowledgment}
This project was supported by the research project FONDECYT N$^o$ 11180881.

\bibliographystyle{plain}
\bibliography{main}

\begin{thebibliography}{10}

\bibitem{case_study_omp_reduce}
Mahwish Arif and Hans Vandierendonck.
\newblock A case study of openmp applied to map/reduce-style computations.
\newblock In Christian Terboven, Bronis~R. de~Supinski, Pablo Reble, Barbara~M.
  Chapman, and Matthias~S. M{\"u}ller, editors, {\em OpenMP: Heterogenous
  Execution and Data Movements}, pages 162--174, Cham, 2015. Springer
  International Publishing.

\bibitem{bedorf2012sparse}
Jeroen B{\'e}dorf, Evghenii Gaburov, and Simon~Portegies Zwart.
\newblock A sparse octree gravitational n-body code that runs entirely on the
  gpu processor.
\newblock {\em Journal of Computational Physics}, 231(7):2825--2839, 2012.

\bibitem{Brent_1974}
Richard~P. Brent.
\newblock The parallel evaluation of general arithmetic expressions.
\newblock {\em J. ACM}, 21(2):201--206, April 1974.

\bibitem{8705253}
R.~{Carrasco}, R.~{Vega}, and C.~A. {Navarro}.
\newblock Analyzing gpu tensor core potential for fast reductions.
\newblock In {\em 2018 37th International Conference of the Chilean Computer
  Science Society (SCCC)}, pages 1--6, Nov 2018.

\bibitem{CARTER2018148}
Francisco Carter, Nancy Hitschfeld, Crist{\'o}bal~A. Navarro, and Rodrigo Soto.
\newblock Gpu parallel simulation algorithm of brownian particles with excluded
  volume using delaunay triangulations.
\newblock {\em Computer Physics Communications}, 229:148 -- 161, 2018.

\bibitem{CERDA20188}
Mauricio Cerda, Crist{\'o}bal~A. Navarro, Juan Silva, Scott~R. Waitukaitis,
  Nicolás Mujica, and Nancy Hitschfeld.
\newblock A high-speed tracking algorithm for dense granular media.
\newblock {\em Computer Physics Communications}, 227:8 -- 16, 2018.

\bibitem{Chaitanya2017}
Chakravarty Reddy~Alla Chaitanya, Anton Kaplanyan, Christoph Schied, Marco
  Salvi, Aaron Lefohn, Derek Nowrouzezahrai, and Timo Aila.
\newblock Interactive reconstruction of noisy monte carlo image sequences using
  a recurrent autoencoder.
\newblock {\em ACM Trans. Graph.}, 36(4), 2017.

\bibitem{Chan:2007:MAA:1250790.1250877}
Timothy~M. Chan.
\newblock More algorithms for all-pairs shortest paths in weighted graphs.
\newblock In {\em Proceedings of the Thirty-ninth Annual ACM Symposium on
  Theory of Computing}, STOC '07, pages 590--598, New York, NY, USA, 2007. ACM.

\bibitem{inproceedings}
Philip Colangelo, Nasibeh Nasiri, Eriko Nurvitadhi, Asit Mishra, Martin
  Margala, and Kevin Nealis.
\newblock Exploration of low numeric precision deep learning inference using
  intel® fpgas.
\newblock pages 73--80, 04 2018.

\bibitem{10.1145/3330345.3331057}
Abdul Dakkak, Cheng Li, Jinjun Xiong, Isaac Gelado, and Wen-mei Hwu.
\newblock Accelerating reduction and scan using tensor core units.
\newblock In {\em Proceedings of the ACM International Conference on
  Supercomputing}, ICS ’19, page 46–57, New York, NY, USA, 2019.
  Association for Computing Machinery.

\bibitem{mapreduce}
Jeffrey Dean and Sanjay Ghemawat.
\newblock Mapreduce: Simplified data processing on large clusters.
\newblock In {\em Proceedings of the 6th Conference on Symposium on Operating
  Systems Design \& Implementation - Volume 6}, OSDI'04, pages 10--10,
  Berkeley, CA, USA, 2004. USENIX Association.

\bibitem{gunther2007realtime}
Johannes Gunther, Stefan Popov, Hans-Peter Seidel, and Philipp Slusallek.
\newblock Realtime ray tracing on gpu with bvh-based packet traversal.
\newblock In {\em 2007 IEEE Symposium on Interactive Ray Tracing}, pages
  113--118. IEEE, 2007.

\bibitem{Haidar:2017}
Azzam Haidar, Panruo Wu, Stanimire Tomov, and Jack Dongarra.
\newblock Investigating half precision arithmetic to accelerate dense linear
  system solvers.
\newblock In {\em Proceedings of the 8th Workshop on Latest Advances in
  Scalable Algorithms for Large-Scale Systems}, ScalA '17, pages 10:1--10:8,
  New York, NY, USA, 2017. ACM.

\bibitem{5325422}
T.~{Hamada}, K.~{Benkrid}, K.~{Nitadori}, and M.~{Taiji}.
\newblock A comparative study on asic, fpgas, gpus and general purpose
  processors in the $o(n^2)$ gravitational n-body simulation.
\newblock In {\em 2009 NASA/ESA Conference on Adaptive Hardware and Systems},
  pages 447--452, July 2009.

\bibitem{harris_2005}
Mark Harris.
\newblock Mapping computational concepts to gpus.
\newblock In {\em ACM SIGGRAPH 2005 Courses}, SIGGRAPH '05, New York, NY, USA,
  2005. ACM.

\bibitem{harris2007optimizing}
Mark Harris.
\newblock Optimizing cuda.
\newblock 2007.

\bibitem{0031-9155-56-22-002}
Xun Jia, Xuejun Gu, Yan~Jiang Graves, Michael Folkerts, and Steve~B Jiang.
\newblock Gpu-based fast monte carlo simulation for radiotherapy dose
  calculation.
\newblock {\em Physics in Medicine \& Biology}, 56(22):7017, 2011.

\bibitem{jia2019dissecting}
Zhe Jia, Marco Maggioni, Jeffrey Smith, and Daniele~Paolo Scarpazza.
\newblock Dissecting the nvidia turing t4 gpu via microbenchmarking.
\newblock {\em arXiv preprint arXiv:1903.07486}, 2019.

\bibitem{DBLP:journals/corr/abs-1804-06826}
Zhe Jia, Marco Maggioni, Benjamin Staiger, and Daniele~Paolo Scarpazza.
\newblock Dissecting the {NVIDIA} volta {GPU} architecture via
  microbenchmarking.
\newblock {\em CoRR}, abs/1804.06826, 2018.

\bibitem{jouppi2018motivation}
Norman Jouppi, Cliff Young, Nishant Patil, and David Patterson.
\newblock Motivation for and evaluation of the first tensor processing unit.
\newblock {\em IEEE Micro}, 38(3):10--19, 2018.

\bibitem{tpu_google_2017}
Norman~P. Jouppi, Cliff Young, Nishant Patil, David Patterson, Gaurav Agrawal,
  Raminder Bajwa, Sarah Bates, and et~al.
\newblock In-datacenter performance analysis of a tensor processing unit.
\newblock {\em SIGARCH Comput. Archit. News}, 45(2):1--12, June 2017.

\bibitem{lecun2015deep}
Yann LeCun, Yoshua Bengio, and Geoffrey Hinton.
\newblock Deep learning.
\newblock {\em nature}, 521(7553):436--444, 2015.

\bibitem{tensor_cores_2018}
Stefano Markidis, Steven Wei~Der Chien, Erwin Laure, Ivy~Bo Peng, and
  Jeffrey~S. Vetter.
\newblock {NVIDIA} tensor core programmability, performance {\&} precision.
\newblock {\em CoRR}, abs/1803.04014, 2018.

\bibitem{martineau2018benchmarking}
Matt Martineau, Patrick Atkinson, and Simon McIntosh-Smith.
\newblock Benchmarking the nvidia v100 gpu and tensor cores.
\newblock In {\em European Conference on Parallel Processing}, pages 444--455.
  Springer, 2018.

\bibitem{8392762}
C.~A. Navarro, M.~Vernier, N.~Hitschfeld, and B.~Bustos.
\newblock Competitiveness of a non-linear block-space gpu thread map for
  simplex domains.
\newblock {\em IEEE Transactions on Parallel and Distributed Systems}, pages
  1--1, 2018.

\bibitem{navarro_hitschfeld-kahler_mateu_2014}
Crist{\'o}bal~A. Navarro, Nancy Hitschfeld-Kahler, and Luis Mateu.
\newblock A survey on parallel computing and its applications in data-parallel
  problems using gpu architectures.
\newblock {\em Communications in Computational Physics}, 15(2):285–329, 2014.

\bibitem{NAVARRO201648}
Crist{\'o}bal~A. Navarro, Wei Huang, and Youjin Deng.
\newblock Adaptive multi-gpu exchange monte carlo for the 3d random field ising
  model.
\newblock {\em Computer Physics Communications}, 205:48 -- 60, 2016.

\bibitem{nickolls_2008}
John Nickolls, Ian Buck, Michael Garland, and Kevin Skadron.
\newblock Scalable parallel programming with cuda.
\newblock In {\em ACM SIGGRAPH 2008 Classes}, SIGGRAPH '08, pages 16:1--16:14,
  New York, NY, USA, 2008. ACM.

\bibitem{CUB}
{Nvidia Corporation}.
\newblock {CUB} library, 2020.

\bibitem{cuda}
{NVIDIA Corporation}.
\newblock {NVIDIA CUDA C} programming guide, 2020.

\bibitem{Putnam:2014:RFA:2665671.2665678}
Andrew Putnam, Adrian~M. Caulfield, Eric~S. Chung, Derek Chiou, Kypros
  Constantinides, John Demme, Hadi Esmaeilzadeh, Jeremy Fowers, Gopi~Prashanth
  Gopal, Jan Gray, Michael Haselman, Scott Hauck, Stephen Heil, Amir Hormati,
  Joo-Young Kim, Sitaram Lanka, James Larus, Eric Peterson, Simon Pope, Aaron
  Smith, Jason Thong, Phillip~Yi Xiao, and Doug Burger.
\newblock A reconfigurable fabric for accelerating large-scale datacenter
  services.
\newblock In {\em Proceeding of the 41st Annual International Symposium on
  Computer Architecuture}, ISCA '14, pages 13--24, Piscataway, NJ, USA, 2014.
  IEEE Press.

\bibitem{omp_ompi}
R.~Rabenseifner, G.~Hager, and G.~Jost.
\newblock Hybrid mpi/openmp parallel programming on clusters of multi-core smp
  nodes.
\newblock In {\em 2009 17th Euromicro International Conference on Parallel,
  Distributed and Network-based Processing}, pages 427--436, Feb 2009.

\bibitem{SCHMIDHUBER201585}
J{\"u}rgen Schmidhuber.
\newblock Deep learning in neural networks: An overview.
\newblock {\em Neural Networks}, 61:85 -- 117, 2015.

\bibitem{stuart2011multi}
Jeff~A Stuart and John~D Owens.
\newblock Multi-gpu mapreduce on gpu clusters.
\newblock In {\em 2011 IEEE International Parallel \& Distributed Processing
  Symposium}, pages 1068--1079. IEEE, 2011.

\bibitem{10.1007/BFb0016236}
Joachim von~zur Gathen.
\newblock Parallel arithmetic computations: A survey.
\newblock In Jozef Gruska, Branislav Rovan, and Juraj Wiedermann, editors, {\em
  Mathematical Foundations of Computer Science 1986}, pages 93--112, Berlin,
  Heidelberg, 1986. Springer Berlin Heidelberg.

\bibitem{wolfram1983statistical}
Stephen Wolfram.
\newblock Statistical mechanics of cellular automata.
\newblock {\em Reviews of modern physics}, 55(3):601, 1983.

\bibitem{Zhao:2017:ABC:3020078.3021741}
Ritchie Zhao, Weinan Song, Wentao Zhang, Tianwei Xing, Jeng-Hau Lin, Mani
  Srivastava, Rajesh Gupta, and Zhiru Zhang.
\newblock Accelerating binarized convolutional neural networks with
  software-programmable fpgas.
\newblock In {\em Proceedings of the 2017 ACM/SIGDA International Symposium on
  Field-Programmable Gate Arrays}, FPGA '17, pages 15--24, New York, NY, USA,
  2017. ACM.

\end{thebibliography}
\vspace{12pt}

\begin{appendices}
\section{CUDA Tensor Core MMA example}
\label{app_mma_example}
\begin{lstlisting}[language=C++, caption=Simple MMA CUDA Kernel, basicstyle=\footnotesize]
#include <mma.h>
#define BSIZE 256 // threads per block
#define WPB 8     // warps per block
#define FS 256    // fragment total size
using namespace nvcuda::wmma;

__global__ void mmakernel(half *a,half *b,float *c){
    // (1) declare fragments
    fragment<matrix_a,16,16,16,half,col_major> af;
    fragment<matrix_b,16,16,16,half,row_major> bf;
    fragment<accumulator,16,16,16,float> cf;

    // (2) C as constant mat
    float my_const = 1.0f;
    fill_fragment(cf, my_const);
    
    // (3) offset for unique regions per warp
    // warp ID
    int WID = threadIdx.x >> 5;
    int off = blockIdx.x*WPB*FS + WID*FS;

    // (4) Load data into A and B fragments
    load_matrix_sync(af, a + off, 16);
    load_matrix_sync(bf, b + off, 16);

    // (5) matrix multiply accumulate
    mma_sync(cf, af, bf, cf);

    // (6) store result 
    store_matrix_sync(c+off,cf,16,mem_row_major);
}
\end{lstlisting}
\section{Results with a TITAN RTX GPU}
\label{app_results_rtx}
\begin{figure*}[ht!]
      \centering
      \begin{subfigure}[b]{0.49\textwidth}
      \includegraphics[scale=.55]{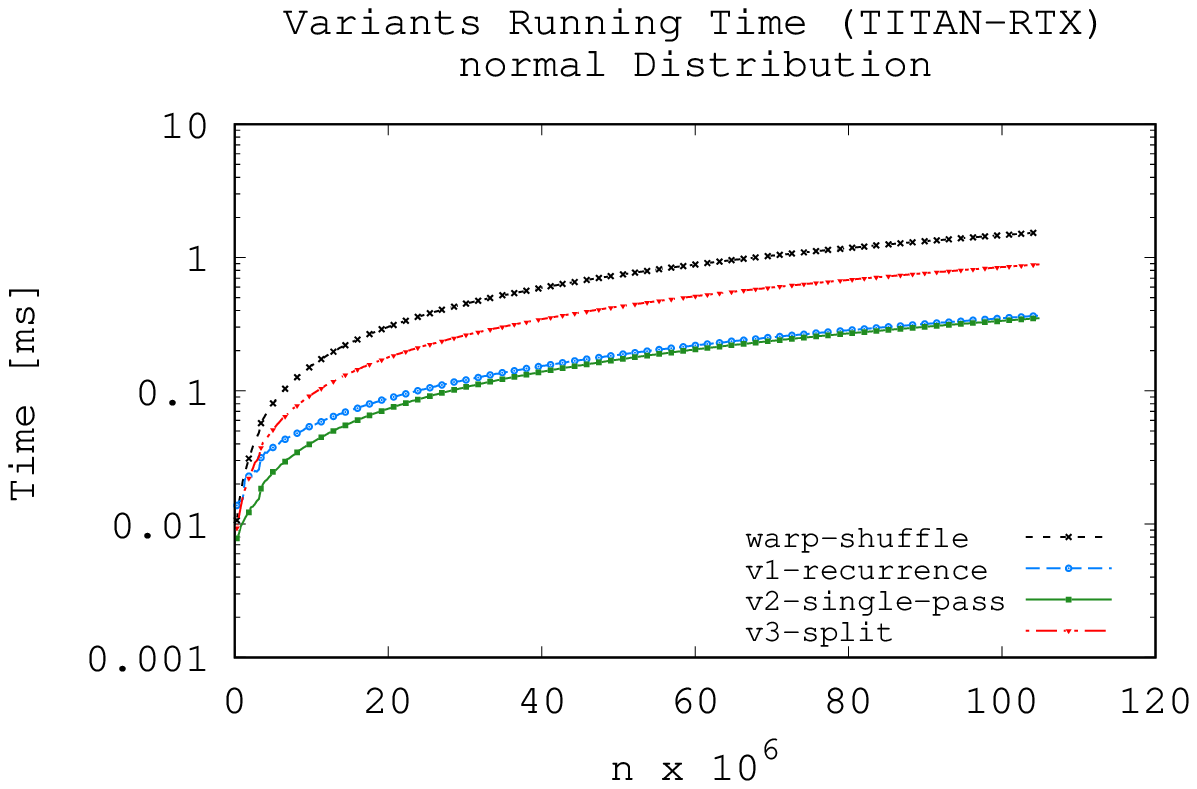}
      \end{subfigure}
      \begin{subfigure}[b]{0.49\textwidth}
      \includegraphics[scale=.55]{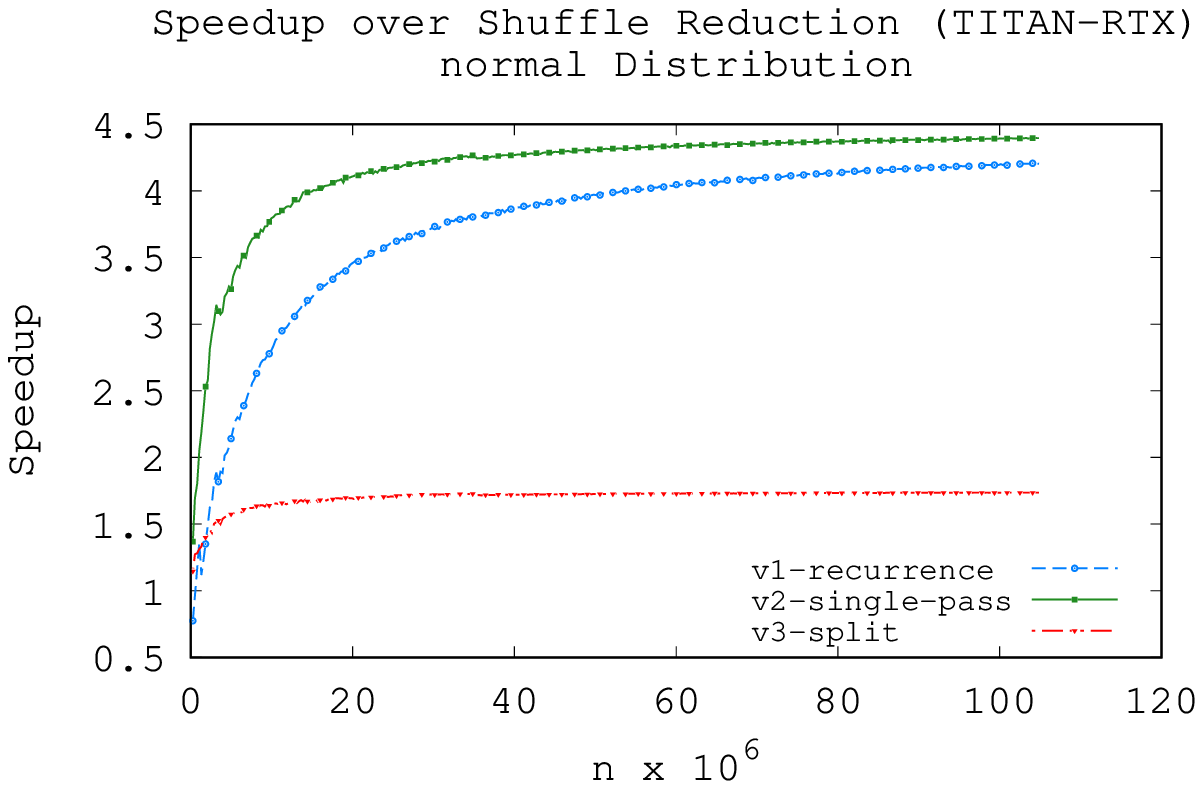}
      \end{subfigure}
      \\
      \begin{subfigure}[b]{0.49\textwidth}
      \includegraphics[scale=.55]{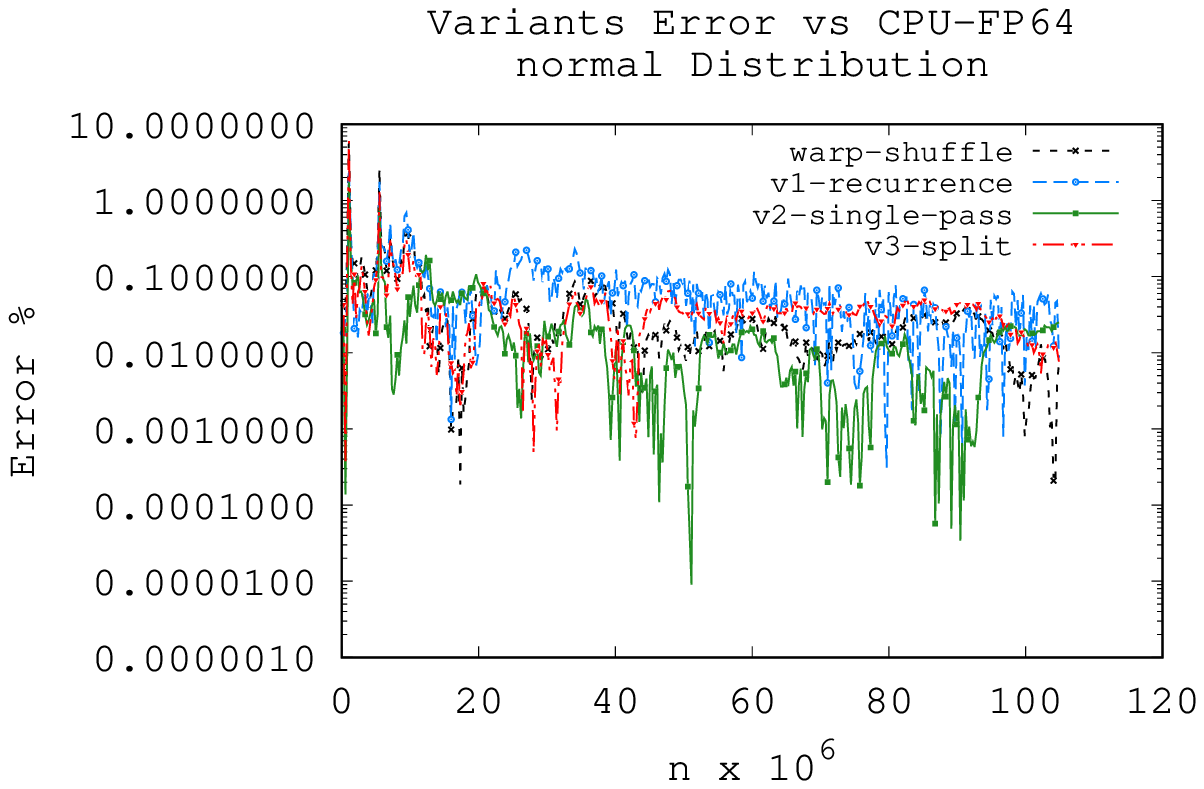}
      \end{subfigure}
      \begin{subfigure}[b]{0.49\textwidth}
      \includegraphics[scale=.55]{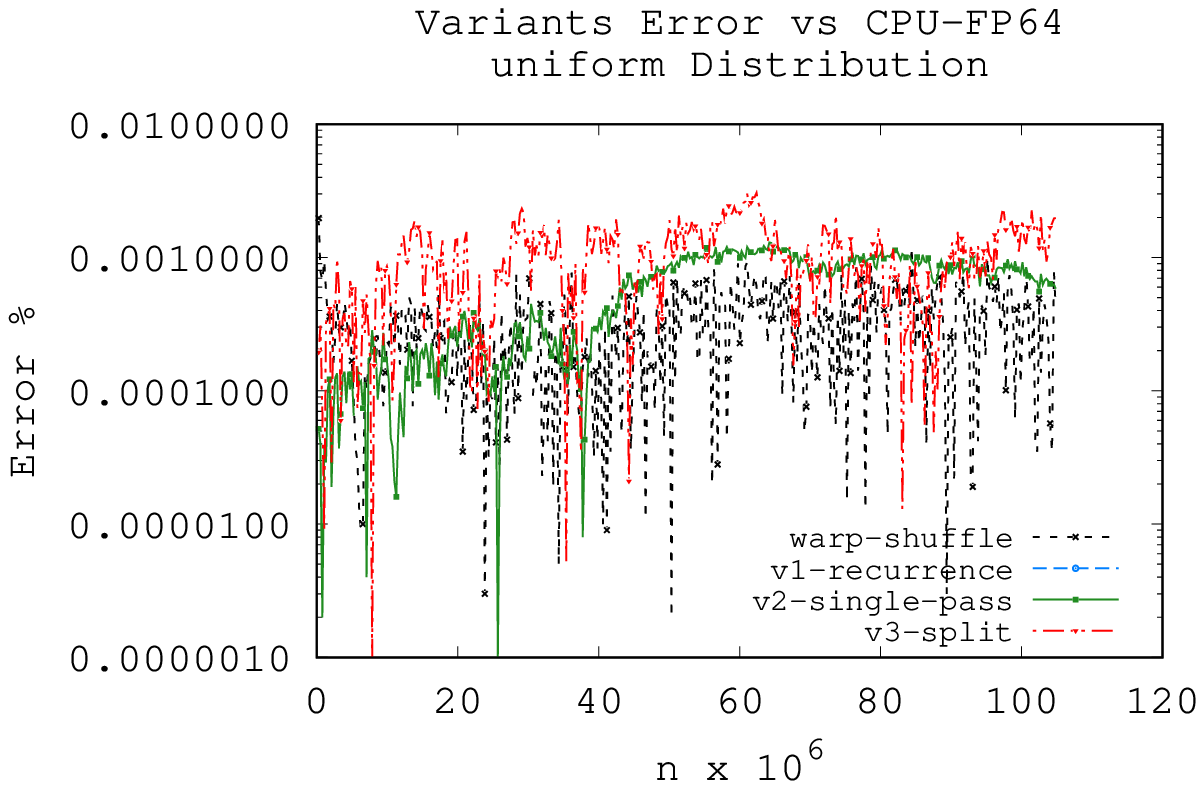}
      \end{subfigure}
      \caption{Runtime, speedup and numerical error of all variants with respect to a warp-shuffle reduction.}
      \label{fig_app_variants_rtx}
\end{figure*}
\begin{figure*}[ht!]
      \centering
      \begin{subfigure}[b]{0.49\textwidth}
      \includegraphics[scale=.55]{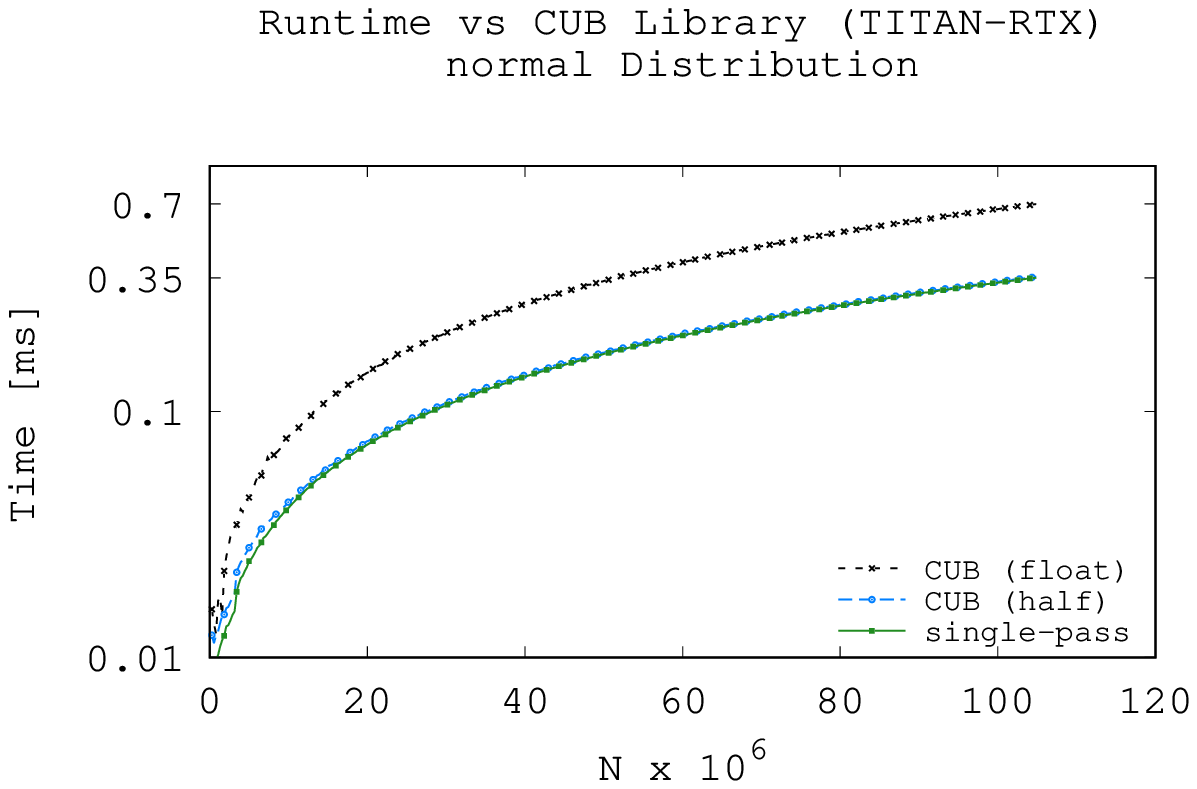}
      \end{subfigure}
      \begin{subfigure}[b]{0.49\textwidth}
      \includegraphics[scale=.55]{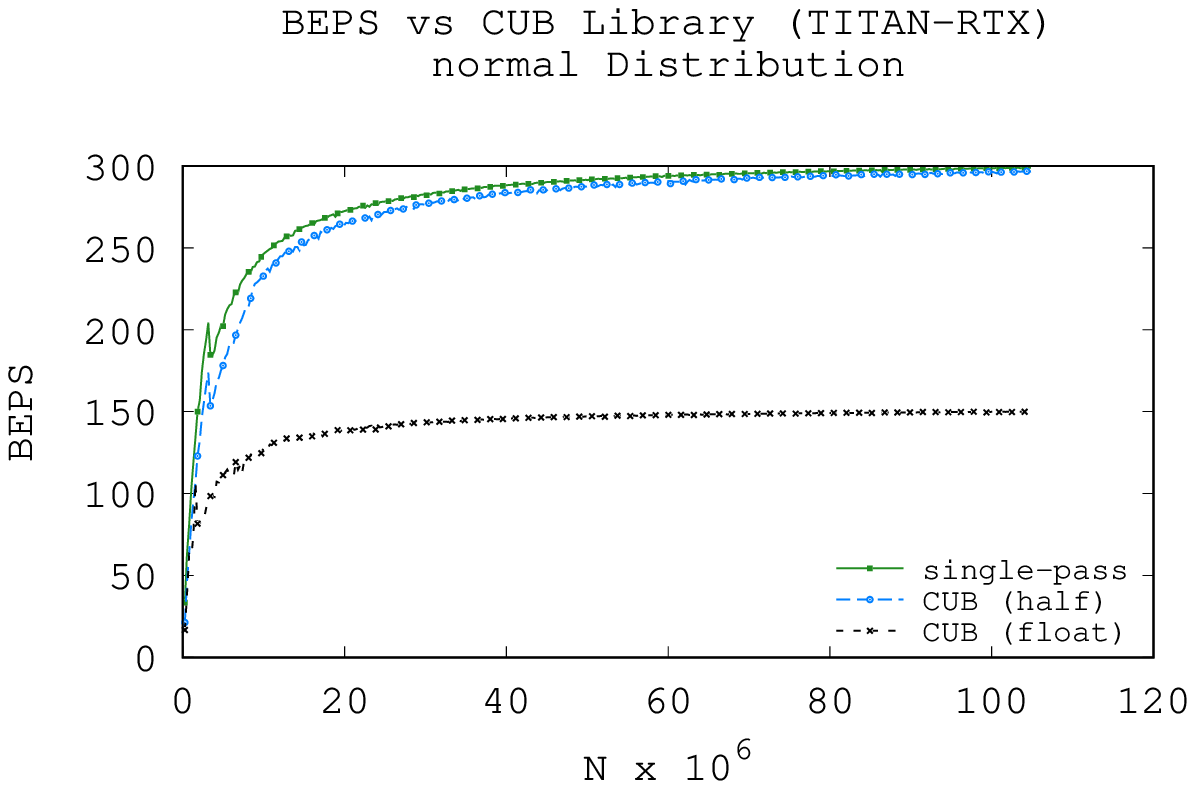}
      \end{subfigure}
      \\
      \begin{subfigure}[b]{0.49\textwidth}
      \includegraphics[scale=.55]{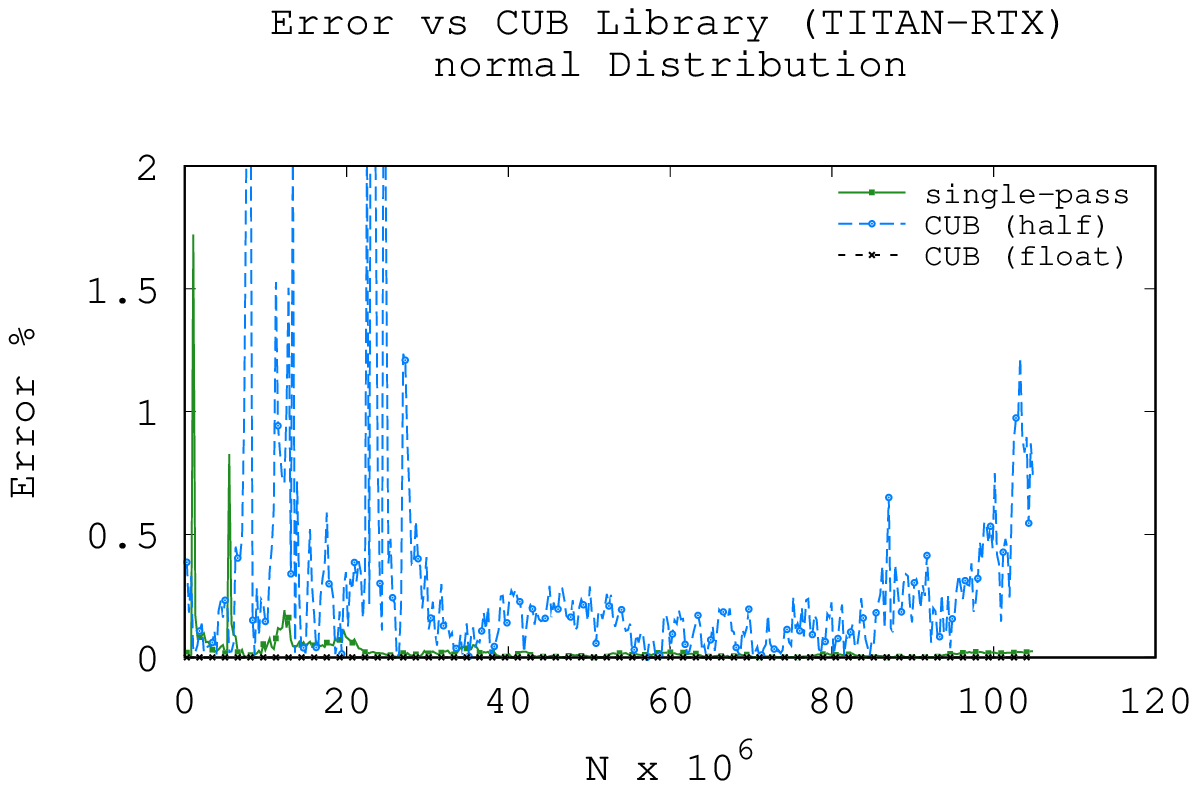}
      \end{subfigure}
      \begin{subfigure}[b]{0.49\textwidth}
      \includegraphics[scale=.55]{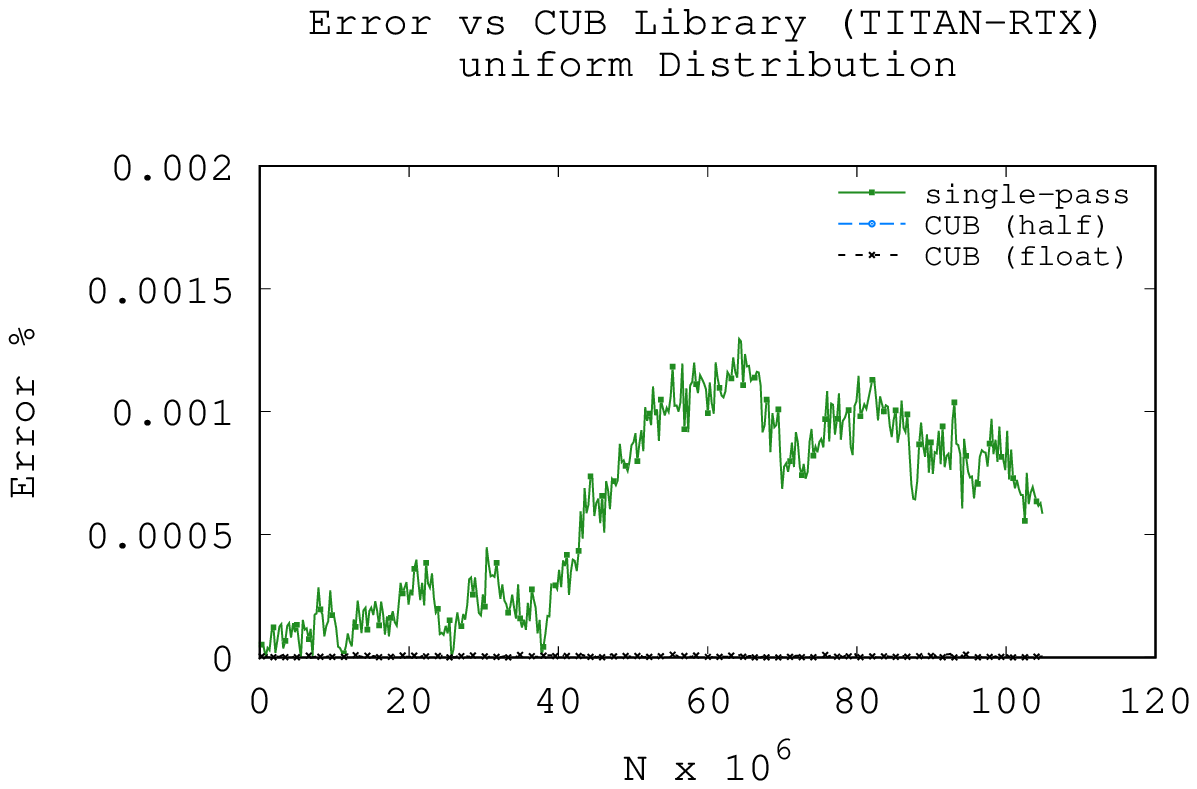}
      \end{subfigure}
      \caption{Runtime, BEPS and numerical error (normal and uniform) of single-pass variant and CUB reductions.}
      \label{fig_app_comparison_rtx}
\end{figure*}
\begin{figure*}[ht!]
      \begin{subfigure}[b]{0.49\textwidth}
      \includegraphics[scale=.55]{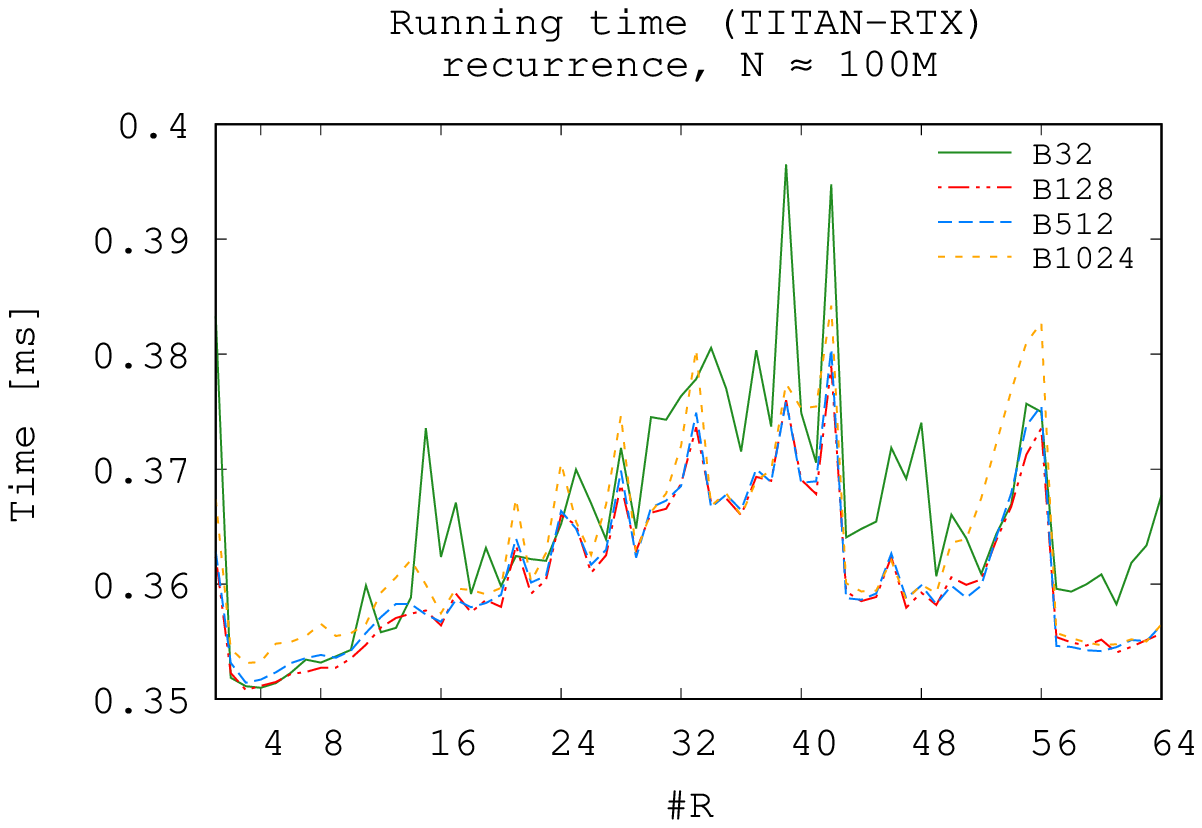}
      \end{subfigure}
      \begin{subfigure}[b]{0.49\textwidth}
      \includegraphics[scale=.55]{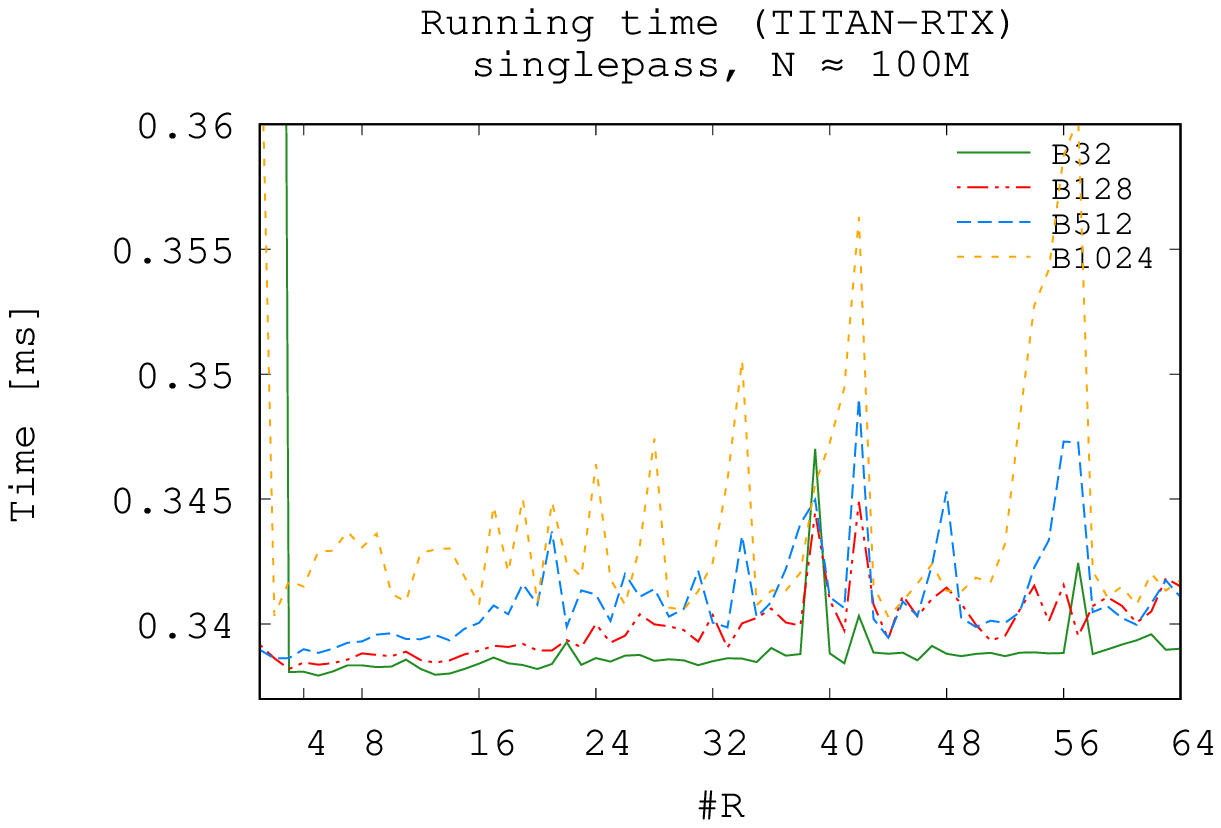}
      \end{subfigure}
      \begin{subfigure}[b]{0.49\textwidth}
      \includegraphics[scale=.55]{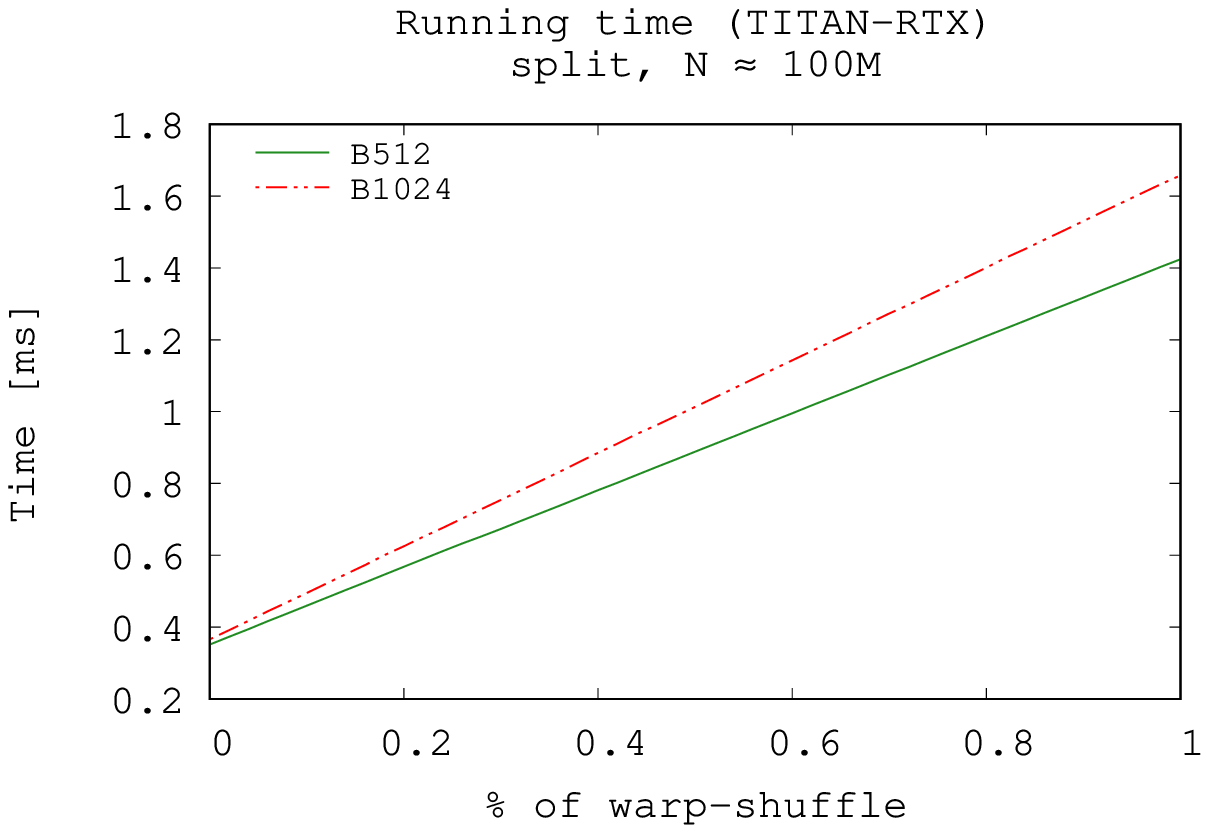}
      \end{subfigure}
      \caption{Plots for configuring $R,B$ values and split percentage values.}
      \label{fig_app_conf_rtx}
\end{figure*}
\end{appendices}
\end{document}